\RequirePackage{fix-cm}
\documentclass[numbook]{svjour3} 
\counterwithin*{section}{part}

\usepackage{url}
\usepackage{relsize}
\usepackage{lscape}
\usepackage{multirow}
\usepackage{ragged2e}
\usepackage{adjustbox, tabularx,booktabs}
\usepackage{dcolumn}
\usepackage{float}
\floatstyle{plaintop}
\usepackage{adjustbox}
\restylefloat{table}
\usepackage{graphicx}
\usepackage[margin=1in]{geometry}

\usepackage{booktabs} 
\usepackage{array} 
\usepackage{paralist} 
\usepackage{verbatim} 
\usepackage{enumitem}
\usepackage{ifthen}
\usepackage{xspace}
\usepackage[table]{xcolor}
\usepackage{fancybox} 
\usepackage{soul}
\usepackage{color, colortbl}
\usepackage{balance}

\usepackage{graphicx}
\usepackage{titlesec}
\usepackage{longtable}
\usepackage{tabularx}
\usepackage[comma,authoryear]{natbib}
\usepackage{amsmath,amssymb,amsfonts}
\usepackage{cleveref}
\usepackage{longtable}
\usepackage{cite}
\usepackage{fancybox} 
\usepackage{amsmath,amssymb,amsfonts}
\usepackage{geometry}
\usepackage{graphicx}
\usepackage{titlesec}
\usepackage{longtable}
\usepackage{algorithm}
\usepackage{algpseudocode}
\usepackage{tabularx}
\usepackage{blindtext}
\usepackage{textcomp}
\usepackage{booktabs}
\usepackage{chngcntr}
\usepackage{framed}
\usepackage{mdframed}
\usepackage{cleveref}
\usepackage{amsmath}
\counterwithout{table}{section}
\counterwithout{figure}{section}
\counterwithout{equation}{section}

\newcommand{\takeaway}[2]{%
    \vspace{-1em}
    \begin{justify}%
        \begin{minipage}{14cm}%
            \begin{mdframed}[
                backgroundcolor=lightgray,
                roundcorner=5pt,
                linewidth=1pt,
                linecolor=black,
                innerbottommargin=10pt,
                innertopmargin=0pt,
                innerleftmargin=10pt,
                innerrightmargin=10pt,
                topline=true,
                bottomline=true,
            ]
                \vspace{5pt} 
                \raggedright\textbf{#1} 
                \vspace{5pt} 
                \noindent\makebox[\linewidth]{\rule{\linewidth}{1pt}}
                #2 
            \end{mdframed}
        \end{minipage}
    \end{justify}
}

\usepackage{url}
\usepackage{relsize}
\usepackage{lscape}
\usepackage{multirow}
\usepackage{ragged2e}
\usepackage{adjustbox, tabularx,booktabs}
\usepackage{dcolumn}
\usepackage{float}
\floatstyle{plaintop}
\usepackage{adjustbox}
\restylefloat{table}
\usepackage{geometry}
\usepackage{graphicx}

\usepackage{booktabs} 
\usepackage{array} 
\usepackage{paralist} 
\usepackage{verbatim} 

\usepackage{ifthen}
\usepackage{xspace}
\usepackage[table]{xcolor}
\usepackage{fancybox} 
\usepackage{soul}
\usepackage{color, colortbl}
\usepackage{balance}

\algnewcommand\algorithmicforeach{\textbf{for each}}
\algdef{S}[FOR]{ForEach}[1]{\algorithmicforeach\ #1\ \algorithmicdo}

\def\base{base model\xspace}
\def\bases{base models\xspace}

\def\llm{PTLM\xspace}
\def\llms{PTLMs\xspace}

\def\ft{Fine-tuned\xspace}
\def\FT{Fine-tuning\xspace}
\def\QT{Quantization\xspace}
\def\qt{Quantized\xspace}

\def\ds{Distilled\xspace}
\def\DP{Deduplication\xspace}
\def\ded{Deduped\xspace}
\def\PR{Pruning\xspace}
\def\PS{Parameter Sharing\xspace}

\def\kds{Knowledge distillation\xspace}

\def\vtype{variant type\xspace}
\def\vtypes{variant types\xspace}

\def\etal{et al. \xspace}

\def\rqa{$RQ_1$ }
\def\rqb{$RQ_2$ }

\def\HF{HF\xspace}

\def\repository{model registry\xspace}

\def\model{downstream model\xspace}

\def\releases{PTLM releases\xspace}

\def\model{PTLM model\xspace}

\def\dmodels{downstream models\xspace}

\def\RQa{What are the current naming and versioning practices of \llms on \HF?}
\def\RQb{What are the \llm variant types, and how are their qualities in terms of reproducibility and transparency on \HF?}
\def\RQc{To what extent do versioning identifiers in \llm names align with actual changes in \llm versions on \HF?}

\begin{document}


\title{Towards Semantic Versioning of Open Pre-trained Language Model Releases on Hugging Face}




\author{Adekunle Ajibode         \and
        Abdul Ali Bangash  
        \and Filipe Roseiro Cogo
        \and Bram Adams
        \and Ahmed E. Hassan}


\institute{Adekunle Ajibode \at
              School of Computing, Queen’s University, Kingston, ON, Canada\\
              \email{ajibode.a@queensu.ca}
           \and
           Abdul Ali Bangash \at
              School of Computing, Queen’s University, Kingston, ON, Canada\\
              \email{abdulali.b@queensu.ca}
           \and
           Filipe R. Cogo \at
               Centre for Software Excellence, Huawei Canada\\
              \email{filipecogo@acm.org}
              \and
           Bram Adams \at
               School of Computing, Queen’s University, Kingston, ON, Canada\\
              \email{bram.adams@queensu.ca}
              \and
           Ahmed E. Hassan \at
               School of Computing, Queen’s University, Kingston, ON, Canada\\
              \email{hassan@queensu.ca}}


\date{Received: date / Accepted: date}

\maketitle

\begin{abstract}
The proliferation of open Pre-trained Language Models (\llms) on model registry platforms like Hugging Face (HF) presents both opportunities and challenges for companies building products around them. Similar to traditional software dependencies, \llms continue to evolve after a release. However, the current state of release practices of \llms on model registry platforms are plagued by a variety of inconsistencies, such as ambiguous naming conventions and inaccessible model training documentation. Given the knowledge gap on current \llm release practices, our empirical study uses a mixed-methods approach to analyze the releases of 52,227 \llms on the most well-known model registry, \HF. Our results reveal 148 different naming practices for \llm releases, with 40.87\% of changes to model weight files not represented in the adopted name-based versioning practice or their documentation. In addition, we identified that the 52,227 \llms are derived from only 299 different \bases (the modified original models used to create 52,227 \llms), with \FT and \QT being the most prevalent modification methods applied to these base models. Significant gaps in release transparency, in terms of training dataset specifications and model card availability, still exist, highlighting the need for standardized documentation. While we identified a model naming practice explicitly differentiating between major and minor \llm releases, we did not find any significant difference in the types of changes that went into either type of releases, suggesting that major/minor version numbers for \llms often are chosen arbitrarily. Our findings provide valuable insights to improve \llm release practices, nudging the field towards more formal semantic versioning practices.

\end{abstract}


\section{Introduction}\label{introduction}
The efficacy of pre-trained language models (\llms) for enhancing various language understanding tasks is widely acknowledged \citep{sarzynska2021detecting}. \llms have initiated a paradigm shift in AI applications, fundamentally altering the landscape of natural language processing (NLP) and catalyzing remarkable progress across diverse software domains. Their success is rooted in their ability to extract patterns from extensive textual datasets, effectively capturing the complexity of human language, thus enabling the development of contextually aware intelligent systems~\citep{wang2022pre}. As such, PTLMs are prominent in popular model registries like Hugging Face (HF) and central to the advancement of AI \citep{zhao2023survey}. They also tend to be larger, better documented and more popular than models from other domains \citep{castano2024analyzing}.

As open-source \llms like the Llama family~\citep{touvron2023llama} have significantly increased in availability and popularity, they have spawned a wide range of \emph{model variants}, each produced through different modification methods, such as fine-tuning, knowledge distillation, pruning, quantization, or any other strategies that alter a model's architecture of training to better suit specific use case. The diversity of these model variants, and the fact that each variant can be further modified into additional variants, poses challenges for stakeholders trying to identify new model versions or variants and understand the associated changes and risks. Given that hundreds of possible variants are continuously evolving, these challenges impact both model developers, who are responsible for creating and maintaining these models, and end users, including industry professionals, practitioners, and academic researchers who rely on the models for various applications. 

In traditional software engineering, the problem of versioning and tracking variants is addressed by the established practice of semantic versioning. It is commonly used by software package management systems and both open-source and commercial software projects to clearly communicate the impact of changes between versions, communicating potential compatibility issues, and reducing integration risks \citep{lam2020putting, decan2019package}. Semantic versioning has been shown to be very useful and important in release engineering, such as helping developers automate dependencies and avoid unnecessary work when components evolve \citep{lam2020putting}, addressing inconsistent breaking changes that impact dependent systems \citep{raemaekers2017semantic}, enabling better version control in continuous delivery environments \citep{carvalho2021deep}, improving compliance with versioning policies across different software ecosystems \citep{decan2019package}, and reducing the impact of breaking changes on client code \citep{ruhroth2014versioning}.

However, popular model registries like \HF  currently lack semantic versioning practices, relying instead on ad hoc naming conventions to communicate updates. These issues are particularly pronounced when analyzing PTLMs, given their popularity and sprawling variants making consistent versioning even more important. To gain a better understanding of how naming conventions and versioning are currently handled within the HF platform, the first and second authors manually explored 50 PTLMs. This selection was random, without prioritizing repositories from top organizations or popular models. While not statistically representative, this exploration provided valuable insights into the prevalent naming and versioning practices on HF, motivating a deeper analysis of these conventions and the need for standardization.

Across these 50 analyzed models, we observed diverse naming practices such as \textit{super-cinnamon/fewshot-followup-multi-e5}\footnote{https://huggingface.co/super-cinnamon/fewshot-followup-multi-e5}, structured with four segments separated by dashes, in contrast to \textit{gsareen07/llama-2-finetune}\footnote{https://huggingface.co/datasets/gsareen07/llama-2-finetune}, which uses three segments to indicate that it is a fine-tuned version of Llama-2. Some models, like \textit{mixedbread-ai/mxbai-rerank-large-v1}\footnote{https://huggingface.co/mixedbread-ai/mxbai-rerank-large-v1}, use version identifiers such as ``v1,'' resembling traditional software practices, while others like \textit{eachadea/ggml-vicuna-7b-1.1}\footnote{https://huggingface.co/eachadea/ggml-vicuna-7b-1.1} do not. We also encountered models uploaded under different names, such as \textit{michellejieli/test\_classifier}\footnote{https://huggingface.co/michellejieli/test\_classifier} and \textit{michellejieli/emotion\_text\_classifier}\footnote{https://huggingface.co/michellejieli/emotion\_text\_classifier}, both using the same base model with nearly identical configurations but lacking dataset specifications. Furthermore, there were variations in the inclusion of model cards and dataset documentation; for instance, \textit{080-ai/flintlock\_3B\_v0.1b}\footnote{https://huggingface.co/080-ai/flintlock\_3B\_v0.1b} did not include a model card but specified the training dataset, whereas \textit{080-ai/tiny-cutlass}\footnote{https://huggingface.co/080-ai/tiny-cutlass} included a model card but omitted details about the training dataset, despite being managed by the same owner. These observations highlight the need for a more thorough empirical exploration of current release practices for \llms on model registries. 

The naming convention of \llms in model stores, and the degree to which they adhere to semantic versioning practices, has not been studied thus far. Prior research has examined a wide variety of release engineering aspect of non-AI systems, including continuous deployment and delivery \citep{shahin2017continuous, bobrovskis2018survey, laukkanen2017problems, kerzazi2016needs}, release notes \citep{abebe2016empirical, bi2020empirical}, release management \citep{michlmayr2007release, khomh2012faster}, and release practices for mobile apps \citep{nayebi2016release, dominguez2019release}. More recently, research has explored ML model and dataset documentation practices \citep{oreamuno2024state, mitchell2019model, wadhwani2020machine, crisan2022interactive, castano2024analyzing}. Yang et al. have analyzed the sub-ecosystem of large language models for code (LLM4Code), focusing on model reuse, documentation practices, and licensing for code-related tasks \citep{yang2024ecosystem}, but did not consider the broader spectrum of \llms used for diverse tasks beyond coding. 
 
To date, there is no empirical research focused specifically on \llm release practices, highlighting the need for more comprehensive studies in this area.

Therefore, this study explores current practices in \llm versioning, the reproducibility of \llms in terms of their provenance and variant types, and the transparency of model cards and dataset documentation on \HF. Specifically, through an empirical analysis of 52,227 \llms on \HF, this paper addresses the following research questions (RQs):

\begin{itemize}
\item[\textbf{$RQ_1.$}] \textbf{\RQa}

\noindent \textit{\underline{Motivation}}: Unclear and inconsistent model naming and versioning conventions can impede the ability of practitioners to effectively understand the communicated changes in model releases. For those developing and publishing models, standardized and meaningful naming practices are crucial for clarifying model identities and tracking modifications. This ensures that changes are documented systematically over time, allowing practitioners to assess whether they can safely update to new model versions.

\noindent \textit{\underline{Findings}}: We found a diverse and heterogeneous landscape of naming practices on \HF, with 148 distinct \llm naming conventions and two types of versioning schemes: major and minor versions. Our analysis also reveals that changes made to model weights are not communicated via the current versioning conventions, indicating a high level of implicit versioning.

\item[\textbf{$RQ_2.$}] \textbf{\RQb}

\noindent \textit{\underline{Motivation}}: Ensuring reproducibility and transparency is crucial for evaluating the reliability and practicality of \llms, as it allows for consistent verification of model performance and fosters trust in the results. Reproducibility ensures that a model can deliver consistent outcomes when retrained or fine-tuned under similar conditions, which is essential for validating scientific claims and practical applications. Transparency, through detailed model card and dataset documentation, provides essential information about training processes and datasets used, enabling users to understand and assess the model’s quality, limitations, and potential biases. By understanding these aspects, practitioners can ensure that \llms are both reliable and trustworthy, facilitating more informed decision-making. 

\noindent \textit{\underline{Findings}}: Since 2022, 299 distinct models, including popular choices like Gemma \citep{team2024gemma}, Mistral \citep{jiang2023mistral}, Llama \citep{touvron2023llama}, and Bert \citep{devlin2018bert},

have served as base models for \llm variant releases. Our manual analysis identified 15 different keywords that can be translated into four different \llm variant types: Fine-tuning, Quantization, Knowledge Distillation, and Deduplication. However, only 17\% of these \llm variant  releases explicitly mentioned keywords corresponding to their variant types, potentially limiting users' ability to accurately reproduce the models and assess their suitability and performance for specific tasks. Furthermore, we observed that only 15.6\%

of \llm releases included training dataset information within their repositories, with even fewer providing details in model cards (12\%) or dataset source links (2\%). This lack of transparency may hinder user understanding and responsible model utilization. Additionally, we noted inconsistencies in model card documentation across different variant types, highlighting the need for standardized documentation practices to enhance transparency in \llm releases on model registry platforms.

\item[\textbf{$RQ_3.$}] \textbf{\RQc} \

\noindent \textit{\underline{Motivation}}: Unlike traditional software engineering, where version numbers typically indicate clear changes such as bug fixes (patch) or feature additions (minor), the specific improvements associated with version updates in \llms~often lack clarity. This ambiguity can hinder practitioners from understanding the nature and impact of updates, making it challenging to decide whether to adopt new versions. By examining how accurately major and minor versioning identifiers in model names reflect the changes observed between model versions, we aim to evaluate the consistency of name-based versioning practices. This assessment is important for determining whether current practices effectively communicate changes and for guiding model owners in refining their versioning strategies, potentially adopting more standardized approaches like semantic versioning.

\noindent \textit{\underline{Findings}}: Major versions exhibit significantly more types of changes, averaging 31 changes, compared to minor versions, which average 10 changes. We grouped these changes into nine categories and observed that the differences between major and minor versions across these categories are not statistically significant. This suggests that practitioners may be using version identifiers in model names arbitrarily, indicating a misalignment between the change types and the identifiers specified in the model names. Additionally, when changes are made to configurations, training libraries, or performance metrics, there is consistently a corresponding change in the performance of the \llms. 
\end{itemize}

Our findings highlight the heterogeneity in model naming conventions, versioning, and release quality within model registry platforms for \llms, emphasizing the need for improved release practices. These improvements should include meaningful and unambiguous model naming, clearer versioning, and comprehensive documentation of datasets, model variant types, and training information through model cards. These measures are important for ensuring the ability to replicate model performance, thereby maintaining the stability and reliability of applications that rely on these models. Specifically, our study provides the following contributions:

\begin{itemize}
    \item We pioneer and provide comprehensive understanding of \llm release practices on \HF, covering \llm naming practices, \llm versioning, and \llm reproducibility and transparency attributes.
    \item We identify sources of naming inconsistencies, missing versions, and documentation gaps, proposing strategies such as standardized naming practices, clear versioning alignment, and improved documentation. 
    \item We offer a dataset and publicly share our extraction code to support empirical research in related fields. These resources aim to facilitate further studies for researchers and model developers~\citep{SAILResearch2024}.
\end{itemize}

This paper is structured as follows. \Cref{background} discuss key concepts such as Pre-Trained Language Models, Model Registries, and Naming and Versioning Conventions in Software Engineering, along with related work. \Cref{methodology} outlines the study setup. \Cref{result} presents the findings of the research questions. \Cref{discussion} covers the study’s discussion and implications, while \Cref{ttv} addresses potential threats to validity. Finally, \Cref{conclusion} summarizes the study and outlines key directions for future research.

\section{Background and Related Work} \label{background}
\subsection{Pre-Trained Language Models (PTLMs)}
\label{subsec:pre-trained-llms}
Pre-trained models are generalist models trained on large-scale datasets to learn broad features that can be adapted to various specific tasks. Unlike simple models like logistic regression, which are trained from scratch for specific tasks, advanced models such as deep neural networks benefit from pre-training on diverse data to develop a strong base of generalized knowledge. This approach, combined with sophisticated architectures and extensive datasets, significantly enhances the model's performance and adaptability across different domains, such as image recognition, speech processing, and natural language processing. The synergy of advanced architectures, large data volumes, and high-quality data, rather than the practice of pre-training alone, has been instrumental in improving these models' capabilities and efficiency compared to training from scratch~\citep{mao2020survey}.

One subset of pre-trained models, \llms, are specifically designed for natural language processing (NLP) tasks. \llms, such as BERT~\citep{devlin2018bert}, GPT~\citep{openai2023gpt}, and RoBERTa~\citep{liu2019roberta}, are trained on extensive text corpora to predict natural language tokens. These models serve as the foundational layer for various NLP applications, significantly enhancing capabilities such as text classification, machine translation, and question answering. For the purposes of this study, we consider \llms with a parameter size of 1 million or more. This threshold is based on findings by Eldan et al.~\citep{eldan2023tinystories}, which demonstrated that language models with 1 million parameters can exhibit reasoning capabilities comparable to larger models.

Following the pre-training phase, practitioners can apply various modification methods (e.g., fine-tuning or quantization) to tailor models for specific applications. In this study, we categorize the modification methods mentioned in the model names\footnote{By ``model name," we refer to the repository name, such as roneneldan/TinyStories-1M, which differs from the base model name, such as BERT.} as distinct variant types. Although a model might have undergone different modification methods that are not mentioned in the names, we use the listed method as the basis for defining our variant types. Numerous types of modification methods exist, a few of which are discussed below:

\begin{itemize}
    \item \FT: Further training pre-trained models on a dataset specific to a task to adapt them to new conditions or improve performance on particular tasks. This may include full fine-tuning or parameter-efficient fine-tuning. For example, fine-tuned models are widely used in tasks like question answering and sentiment analysis \citep{min2017question, severyn2015unitn}. 
    \item \DP: Identifying and removing redundant data in datasets before training to improve the quality of training data and prevent model overfitting \citep{kandpal2022deduplicating}.
    \item \kds: Transferring knowledge from a larger teacher model to a smaller student model, reducing model size for deployment on devices with limited resources \citep{sun2019patient}.
    \item \QT: Reducing the model's precision to save space and computational resources, commonly converting model parameters to 8-bit integers for faster computations \citep{jacob2018quantization}.
    \item \PR: Removing less important weights from the models, which reduces the model size and computational cost \citep{zhu2017prune}.
    \item \PS: The technique of keeping the majority of a pre-trained model's parameters fixed while introducing a small number of additional parameters specific to each new task \citep{houlsby2019parameter}. This differs from fine-tuning in that it involves minimal updates to the pre-trained model parameters, focusing instead on leveraging a shared base with specific extensions.
\end{itemize}

Numerous pre-trained models have been released via \HF. Examples include Llama-2 \citep{touvron2023llama}, xlm-roberta-base \citep{conneau2019unsupervised}, and bert-base-uncased \citep{devlin2018bert}. These models are designed for general-purpose NLP tasks and have been widely adopted across various applications. Moreover, these models serve as the foundation for their variant models, which result from the application of one of such modification methods. Examples of these variants include \textit{starmpcc/Asclepius-13B}\footnote{https://huggingface.co/starmpcc/Asclepius-13B}, \textit{starmpcc/Asclepius-7B}\footnote{https://huggingface.co/starmpcc/Asclepius-7B}, and \textit{THUDM/agentlm-13b}\footnote{https://huggingface.co/THUDM/agentlm-13b}.

Therefore, when we mention \bases, we refer to pre-trained models that have not undergone any of the modification methods mentioned above and still retain their original parameters, such as Llama, Mistral, Bert, and other \llms. Variant types refer to the various modification methods, such as fine-tuning or deduplication, with their resulting models being referred to as fine-tuned models or deduplicated models. When we refer to \llms, we mean the models in general, whether they are the original base models or modified variants.

\subsection{Model Registries}
\label{subsec:registry}

Model registries are centralized repositories designed to store, manage, and distribute ML models \citep{xiu2020exploratory}. They serve as essential infrastructure to ensure reproducibility, sharing, and deployment of models across various environments. These registries enable developers to access a wide array of \llms, facilitating the reuse and adaptation of existing models to new problems. Several model registries are available, such as \textit{HF}\footnote{https://huggingface.co/models}, \textit{ONNX}\footnote{https://github.com/onnx/models}, \textit{PyTorch Hub}\footnote{https://pytorch.org/hub/}, \textit{Model-Zoo}\footnote{https://modelzoo.co/}, and \textit{Modelhub}\footnote{http://app.modelhub.ai/}. Among these, \HF stands out as the largest model registry~\citep{jiang2022empirical}, not only because of the volume of hosted models but also due to its comprehensive set of resources, such as inference APIs, model card support, and extensive documentation for managing models.

On \HF, models are systematically organized by their owners, with each release housed in its own repository. Model owners often maintain multiple repositories that include not only models but also associated datasets and spaces for specific tasks. Base models are frequently adapted into new variants or versions, allowing continuous evolution and flexibility in addressing diverse applications. This organization and the extensive resources available on \HF are particularly significant given the platform’s substantial growth, from 500,000 requests per month in May 2021~\citep{kirk2021bias} to over 7 million per month as of now~\citep{jiang2023empirical}. This surge highlights the importance of using \HF as a case study for understanding \llm release practices.

In contrast to similar stores for mobile apps, Linux distribution packages, or programming language libraries, there is no official versioning mechanism for \llms on \HF. Typically, versioning is managed through arbitrary naming schemas and heuristics rather than standardized or conventional systems. Semantic versioning, commonly used in traditional software development, involves assigning version numbers with a structure like major.minor.patch (e.g., 1.0.0), to indicate the level of changes and compatibility~\citep{lam2020putting}. However, such a well-defined versioning system does not exist for \llms on \HF or other model registries mentioned above. The closest existing system for \llm versioning on \HF is based on naming schemas. These naming schemas often encode specific details such as the model's architecture or type, the dataset or task it was trained on, and additional characteristics like model size and version. Although a structured versioning system is essential to support efficient model discovery and usage within the \HF model registry, there is no standard, and arbitrary naming schemas have limitations. Additionally, there are no checks on names, making them prone to typos and inconsistencies, which cannot be enforced. Therefore, while naming schemas practices aim to enhance transparency and reproducibility, they may vary significantly between different models and practitioners.

For instance, model \textit{cross-encoder/ms-marco-MiniLM-L-6-v2}\footnote{https://huggingface.co/cross-encoder/ms-marco-MiniLM-L-6-v2} exemplifies a naming practice where ``cross-encoder'' specifies the model's owner, ``ms-marco'' denotes the associated dataset or task, and ``MiniLM-L-6-v2'' specifies the model's size, number of layers, and version. In contrast, other models such as \textit{michellejieli/NSFW\_text\_classifier}\footnote{https://huggingface.co/michellejieli/NSFW\_text\_classifier} may provide more generalized descriptions without specific architectural or versioning details. Variations in versioning practices can also be observed, such as the use of whole numbers (\textit{cross-encoder/ms-marco-MiniLM-L-6-v2}) versus decimal numbers (\textit{Vezora/Mistral-22B-v0.2}\footnote{https://huggingface.co/Vezora/Mistral-22B-v0.2}) to denote different model updates. These practices influence how models are updated and managed within the \HF ecosystem, impacting their applicability across different downstream tasks.

\begin{figure*}[t]
\centering
\includegraphics[width=\textwidth]{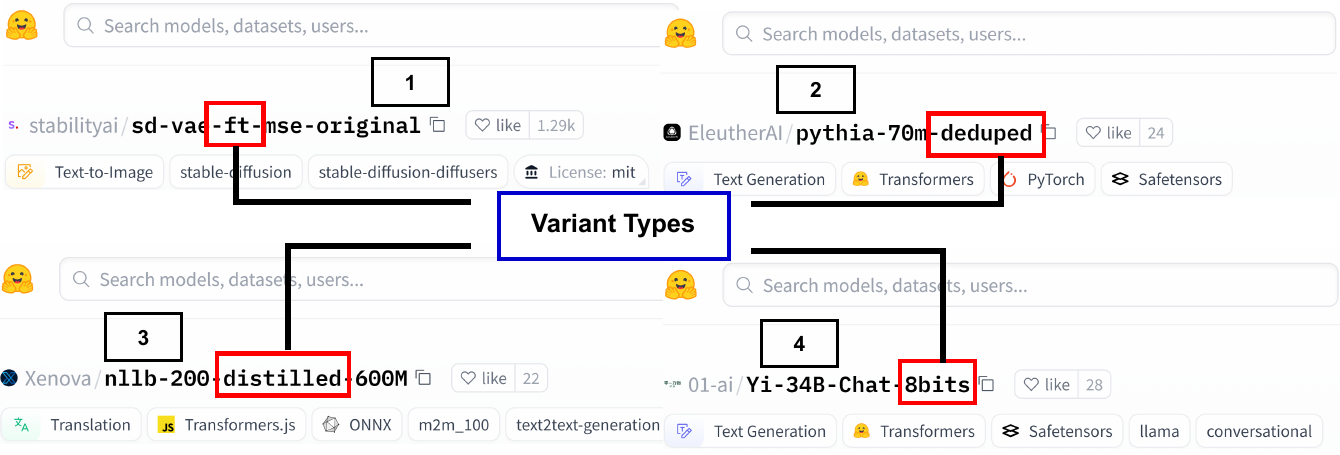}
\caption{Four different examples of how the model modification methods (\vtype) are specified in the model names on \HF repository.}
\label{variant_type_background}
\end{figure*}

\Cref{variant_type_background} illustrates how variant types are often specified in the model names on the \HF repository. For example: Model names might include keywords like ``ft'' or ``fine-tuned'' to indicate that the model has undergone fine-tuning. Keywords like ``deduped'' are used to signify that the deduplication method was applied. Models might be labeled with terms like ``distilled'' to indicate the application of knowledge distillation. Keywords such as ``8bit'' are used to denote quantized models.

Understanding these naming and versioning practices, along with their inconsistencies and limitations, is essential for both model developers and users to effectively navigate and utilize model registry platforms. By elucidating these conventions, this study aims to contribute to improving the transparency and reproducibility of model releases on such platforms.

\subsection{Naming and Versioning Conventions in Software Engineering}
\label{subsec:convention_software} 

Effective naming of software components is important for code readability, maintainability, and collaboration \citep{seacord1998agora, lawrie2007effective, gresta2021naming}, as it simplifies the process of searching for and selecting components for reuse. However, this should not be confused with version naming, which specifically refers to conventions like semantic versioning used to specify version numbers and track changes over time. 

\begin{figure*}[t]
  \centering
  \includegraphics[width=.7\textwidth]{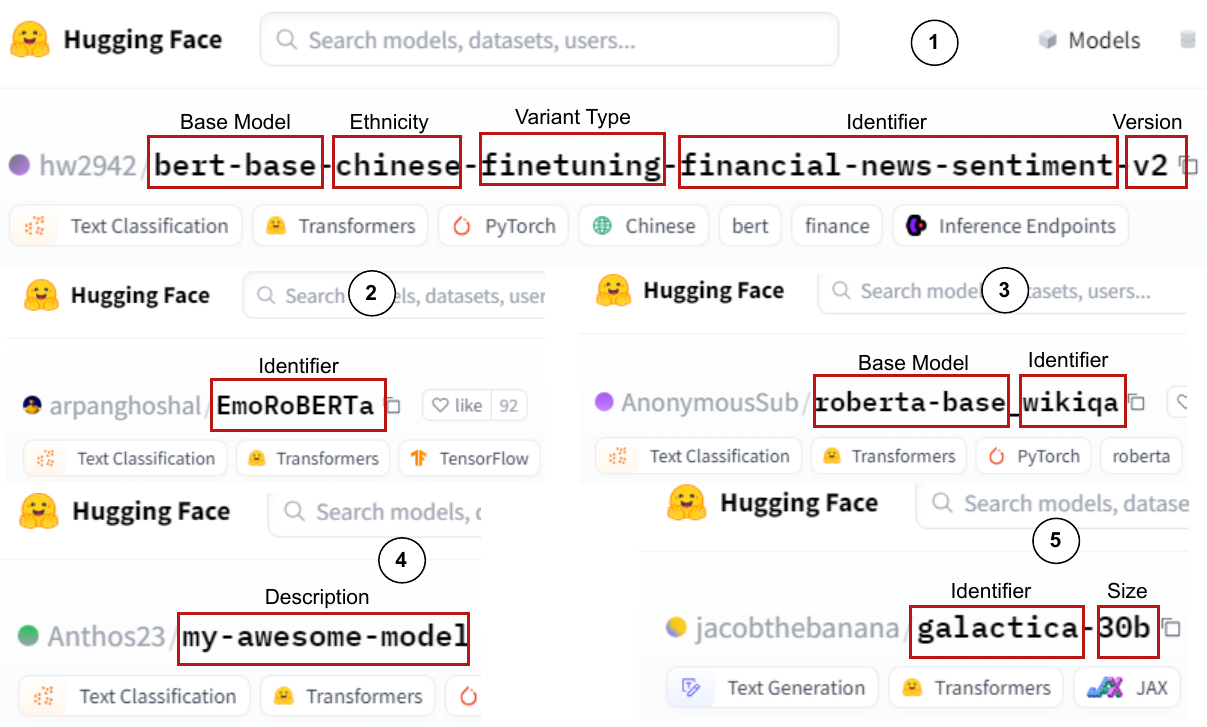}
  \caption{Five different examples of model naming practices on \HF. Some models have 5 segments, while some have less than 2 segments. Each of these examples indicates different information in the names, such as \bases, \vtypes, version, and size.}
  \label{inconsistencies}
\end{figure*}

In contrast, the naming conventions for models on platforms like \HF often involve segmented names with different parts separated by hyphens (-). These segments may represent various attributes of the model, such as \bases, \vtypes, versions, and sizes, as illustrated in \Cref{inconsistencies}. This approach can lead to confusion, as it mixes component naming with version information within the same string.

The challenge with this segmented naming method is that it can overload the naming scheme with versioning details, making it less sustainable. Model names are subject to change, and errors or inconsistencies can occur, complicating the tracking of updates and changes. Proper versioning, using dedicated practices like semantic versioning, is crucial for \llms to ensure accurate tracking of changes and maintain compatibility across versions over time.

In software engineering, the most commonly used versioning mechanism involves incrementing version numbers to indicate changes, with major version increments for backward-incompatible changes, and minor or patch increments for enhancements and bug fixes \citep{stuckenholz2005component, novakouski2012best}, respectively. Minor version increments typically introduce new features or improvements that are backward-compatible, while patch version increments address bug fixes and small changes that do not affect the software’s functionality or compatibility.

\subsection{Related Work}
\label{subsec:related_work}
Numerous studies have extensively investigated identifier naming within software engineering. For instance, Gresta et al. explored naming practices across 40 open-source Java projects, identifying eight distinct approaches \citep{gresta2021naming}. Loomes et al. explored the implications of naming conventions on software maintenance and evolvability, highlighting how traditional naming practices may not fully address the unique needs of software systems \citep{loomes2005naming}. Wenxin et al. investigated naming practices for pre-trained models (PTMs), noting discrepancies in PTM naming on \HF, revealing misalignments with traditional practices, and introducing a novel technique for detecting naming anomalies \citep{jiang2023exploring}. In contrast to these studies, this paper focuses on the naming practices specific to \llms on \HF.

Furthermore, while Jiang et al. provided (amongst others) valuable insights into practitioners' usage of model naming segments \citep{jiang2023exploring} on HF, their work did not analyze this in the context of semantic versioning of models, which is the core focus of our work.

Notably, while Jiang et al. examined individual naming segments, they did not analyze how these segments combine into structured naming conventions, as we do. Similarly, what Jiang et al. regarded as naming conventions in their study actually corresponds to naming segments in our study, since they focused on individual components within the name segment, whereas naming conventions in our study encompass the combination of various segment types.

As a result, we uncovered 12 distinct segments used across 148 unique naming conventions, while Jiang et al. identified 12 distinct segments only without exploring their combinations. Unlike Jiang et al.’s study, which separated aspects such as training process, number of layers, and number of parameters into distinct segments, we grouped these under a single ``training mechanism" segment type, providing a more streamlined categorization. Additionally, their broad ``other" category is undefined, whereas our explicit identification of segments such as ``identifier," ``description," and ``creation date" adds clarity to the taxonomy of name segment types.

Another difference with the study of Jiang et al. is that they did not focus on possible correlations between naming conventions and model popularity (download rate), while our analysis fills this gap, providing actionable insights into the relationship between naming strategies and user engagement.

Furthermore, our findings reveal that only 6.64\% of models on HF include explicit versioning information, with major versioning being the predominant practice, while the aspect of versioning and repository management is absent in Jiang et al.'s study. We also identify significant repository evolution, including frequent updates to model weight files, often without corresponding changes in version identifiers.

In addition, Jiang et al. did not address transparency in metadata or documentation completeness, both of which are vital for effective model reuse. Our analysis highlights that 33\% of PTLMs lack model card documentation, and most fail to provide explicit training dataset metadata. By shedding light on these gaps, our work highlights the need for improved transparency and documentation practices within the HF ecosystem—areas overlooked in Jiang et al.'s investigation.

Lastly, Jiang et al. also trained a model to automatically investigate anomalies in model names in terms of discrepancies between the base model specified in the model name and the one specified in the Hugging Face configuration file. Their results show that their tool effectively detects naming anomalies based solely on architectural information with 92.18\% accuracy—an area we did not cover in our study.

Castano \etal conducted a comprehensive study on ML model evolution and maintenance on the \HF platform, revealing dynamic shifts in model development practices and emphasizing the critical role of systematic maintenance and incremental improvements for long-term model efficacy \citep{castano2024analyzing}. Kathikar \etal examined 110,000 HF model repositories on GitHub, employing static analysis to detect vulnerabilities. They found a significant number of vulnerabilities, with a higher concentration of high-severity issues in popular foundational repositories like Transformers, highlighting the complexity of securing ML models \citep{kathikar2023assessing}. Castano \etal analyzed approximately 170,000 models to investigate HF's environmental sustainability impact. They found that only a small fraction of models reported carbon emissions from training, primarily those trained on HF's infrastructure, which automatically reports emissions. Over time, the percentage of models reporting emissions decreased, but among those that did report, average emissions slightly decreased. The study also identified factors associated with higher carbon emissions \citep{castano2023exploring}. In contrast, our study explores the reproducibility and transparency of \llm releases in \HF, focusing on the consistency and naming practices. 

Similarly, in software and artifact versioning, Novakouski \etal described the challenges of software versioning in service-oriented architectures (SOA) and provided industry guidelines for managing change, emphasizing the impact of versioning on the software life cycle and the importance of a comprehensive versioning policy \citep{novakouski2012best}. Paez proposed a version control strategy for managing new artifacts introduced by DevOps practices, covering artifact identification, versioning tools, naming practices, and traceability, and validated the strategy in three real-world projects \citep{paez2018versioning}. In contrast, our work focuses on the versioning conventions of pre-trained \llms.

In the same vein, \citep{oreamuno2024state} have shown that only a fraction of models and datasets on \HF are properly documented, revealing inconsistencies in ethics and transparency-related information. Mitchell \etal, addressing the need for transparent model reporting, proposed a framework called model cards, advocating for detailed documentation of performance characteristics across various conditions to promote responsible and informed usage of ML models \citep{mitchell2019model}. Toma \etal, focusing on dataset and model management in ML applications, found that most are stored in file systems, lack proper version control integration, and are infrequently updated, leading to issues with availability, traceability, and reproducibility \citep{toma2024exploratory}. Gong \etal, in a comprehensive review of dataset quality in ML, provided valuable guidance for improving the accuracy and efficiency of ML models \citep{gong2023survey}. Lastly, Jiang \etal addressed the scarcity of structured datasets documenting pre-trained model  supply chains by presenting the PeaTMOSS dataset, enabling comprehensive analysis and understanding of pre-trained model adoption and reuse dynamics \citep{jiang2024peatmoss}. 

In terms of model documentation practices, \citep{oreamuno2024state} shed light on the documentation shortcomings of all models on \HF, examining a total of 55,280 models. At the time of their study, the number of models in the model registry was smaller compared to the current number, which has since grown significantly due to influx and shifting community interests. Our research explores the release documentation practices of 196,211 models, focusing specifically on the availability of model cards and training datasets, with a threshold for parameter size. Furthermore, Toma \etal~\citep{toma2024exploratory} studied dataset storage locations and versioning for general ML models on GitHub, highlighting issues with storage and version control integration. However, they did not explore the availability of training datasets for  \llms, which is crucial for reproducibility. In contrast to their work, our research focuses on the availability of training datasets and the versioning of \llms on \HF, specifically examining these critical aspects for models with large parameter sizes.

In addition to the work within the Software Engineering domain, recent studies in related fields have emphasized key aspects of transparency, reproducibility, and documentation in AI and machine learning. \citep{simbeck2022facct} examines legislative approaches to regulating AI in the public sector, focusing on fairness and transparency to address risks associated with AI deployment. \citep{turri2024transparency} explore transparency needs in AI decision-support tools through a comprehensive case study, highlighting the challenges and diverse requirements of end-users. \citep{kinahan2024achieving} tackle reproducibility in EEG machine learning research by introducing the EEG ML Model Card, a standardized documentation tool aimed at addressing issues such as data leakage and flawed model selection. \citep{ahn2024impact} investigate user trust in AI, showing how factors like interpretability and outcome feedback influence trust and task performance, with implications for improving transparency in AI-human collaboration. Finally, \citep{alcobacca2020mfe} address reproducibility in meta-learning by proposing Meta-Feature Extractor (MFE) packages that standardize meta-feature extraction, enabling reliable experimentation.
While these studies generally addressed transparency and reproducibility in various domains, our research examines the transparency and reproducibility of pre-trained language models on Hugging Face, including model cards, dataset documentation, and adaptation methods. By focusing on Hugging Face, we aim to identify gaps that hinder model adoption and integration, providing insights to improve PTLMs accessibility and reliability for both practitioners and users.

\section{Study Setup}\label{methodology}
\begin{figure*}[t]
\centering
\includegraphics[width=\textwidth]{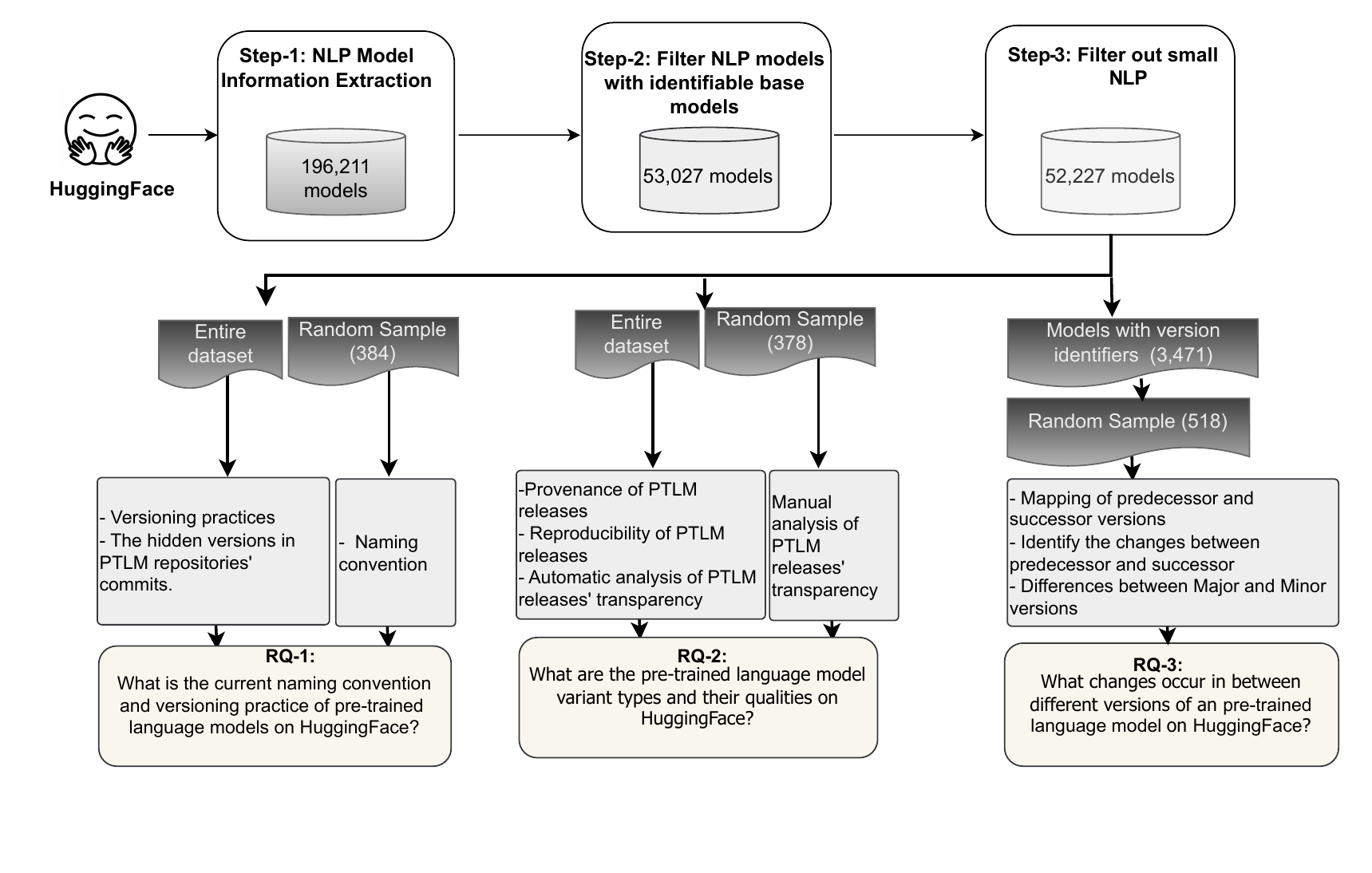}
\caption{Data collection procedure}
\label{framework}
\end{figure*}

This section presents the design of our empirical study addressing the three research questions outlined in the introduction. \Cref{framework} illustrates the procedures we followed to extract and refine our dataset for this research.

\subsection{Data Collection Procedure} \HF contains tens of thousands of \dmodels for dozens of tasks. We outline the three steps that we follow in order to obtain our dataset from the model registry:

\begin{itemize}
   \item \textbf{Step 1: NLP Model Information Extraction:} We conducted a comprehensive extraction of NLP models from \HF, facilitated by the HfApi Client\footnote{https://huggingface.co/docs/huggingface\_hub/package\_reference/hf\_api}. 
   Following a structured procedure for data collection, as depicted in \Cref{framework}, we obtained a total of 196,211 NLP models extracted as of March 17, 2024. 
   \item \textbf{Step 2: Filter NLP models without identifiable base models:} In this step, we aim to focus on models that clearly indicate their base models, as these models typically inherit the characteristics of their base models, including parameter sizes and generative capabilities. Since this step is not straightforward, we explain our approach in detail in the next section. During this process, we filtered out 143,184 (72.97\% of 196,211) models, resulting in 53,027 models. This step was necessary to ensure that we focus our analysis on models with traceable lineage, which are more likely to be robust, mature, and widely used. By doing so, we minimize the likelihood of including experimental or toy models that may lack broader applicability or relevance.
   However, it is important to note that certain base models, such as Llama2, Gemma, and Mistral, are base models themselves and do not have other base models.  These base models are not automatically accessible through the HfAPI we used, and access to them requires manual authentication, which is impractical given the large number of inaccessible models. As a result, 1,497 (1\% of 143,184) models requiring manual authentication were excluded from consideration, as access to them depends on the repository owner’s approval. Instead, for practicality, we focused on open and readily available models, allowing for a more seamless and efficient filtering process. As a result, after Step 2, we retained 53,027 variants that exhibit the characteristics of these base models and their accessible configurations.

    \item \textbf{Step 3: Filter out small NLP}: We filtered out NLP models that either did not have their model size indicated in their safetensors (a recent format designed for efficiently and safely storing model weights\footnote{https://huggingface.co/docs/safetensors/index}) or in their model name. To extract model sizes from the safetensor format, we used the HfAPI Client, while to extract model sizes from model names (e.g., models like meta-llama/Llama-2-7b\footnote{https://huggingface.co/meta-llama/Llama-2-7b}, where ``7b" represents 7 billion parameters), we developed a Python script to gather this information. At the end of this filtering process, 800 models—representing 1.50\% of the 53,026 models—did not have their sizes indicated in either the safetensor format or their model names. Subsequently, we applied a threshold of 1 million parameters to the remaining 52,227 models. All of these turned out to satisfy the size criterion, ensuring that they are sufficiently large, robust, and widely used \citep{eldan2023tinystories}. This constitutes our final dataset for further analysis.

    Note that 14\% of the 52,227 models  do not specify safetensor as their format, suggesting that our dataset may include models from various frameworks such as PyTorch, TensorFlow, and ONNX. However, the absence of a safetensor format specification does not necessarily indicate that a model belongs to one of these other frameworks. We have made this dataset available in our replication package \citep{SAILResearch2024}.
  
\end{itemize}

It is important to note that the difference between the random samples of \rqa and \rqb in \Cref{framework} is due to \rqa being sampled from the entire dataset, whereas \rqb is sampled from the remaining dataset after conducting automatic analysis for that research question.

\subsection{Identifying \bases}\label{identify_base}
We use two distinct approaches to extract the \bases for each \llm. First, we examine the ``model\_type" field in the \HF model configuration file to extract the associated values. This configuration file, in JSON format, contains metadata and parameters defining the model's architecture and behavior. Model owners typically specify their \bases in this field. We then developed a Python script~\citep{SAILResearch2024} to automate the extraction of base model information from the ``model\_type" field. To ensure the accuracy of the extracted base models, we randomly selected 50 samples from the dataset for manual verification. This manual cross-check confirms that the extracted values accurately represent the base models of the studied \llms.

Second, for \llm repositories that did not include configuration files but listed sizes and the \base in the model name, we developed another Python script~\citep{SAILResearch2024} to extract \bases directly from the model names. This script uses the known extracted base models from the ``model\_type'' field as a reference. It first splits the model names into two parts using the standard \HF separator (owner/identifier). Then, it replaces underscores (``\_'') in the identifier with hyphens (``-'') to standardize the format. Finally, it decomposes each model name into its constituent parts, or segments. By comparing these segments with the list of previously retrieved \bases, the script identifies \bases for \llms lacking configuration files.
To ensure the accuracy of the extracted base models, we manually verified 50 of the models for which base models were automatically identified. This cross-check confirmed that the extracted values accurately represent the base models of the studied \llms. Additionally, we randomly selected another 50 models that did not have base models identified through our two approaches for manual verification. This process ensured that these models genuinely lack base models.

It is worth noting that, for the purposes of this study, we define a base model strictly as the root model of an entire model lineage tree, as referenced in the (1) ``model\_type" field within the config JSON file on \HF, rather than the direct parent specified in the (2) ``name\_or\_path" field in the same file. This distinction is necessary because the child model referenced in the latter can itself serve as the parent model to another variant, forming a multi-layered supply chain that complicates the process of tracing a model’s lineage. Our standardized approach ensures consistency across our analysis and provides a clear framework for identifying base models. Future research could expand on this by fully mapping hierarchical relationships between models and their roots, offering deeper insights into the multi-layered structure of model lineages and the evolution of PTLMs over time.

\section{Results}\label{result}

\subsection{\textbf{RQ$_1$:} \RQa} 

\subsubsection{Motivation} Unclear and incoherent model naming and versioning conventions can hinder users' ability to select and utilize the most suitable \llms for their needs. Standardized and meaningful naming practices serve to clarify model identities and streamline the search process. Effective versioning is important, not only for tracking changes and ensuring model reproducibility but also for understanding whether a model update is compatible with current products or if it will cause integration issues. This helps maintain the stability and reliability of systems that rely on these models.

\subsubsection{Approach} 
\emph{Identifying the naming practices of \llms on \HF.} To understand the existing model naming practices on \HF, we employed a manual analysis approach, combining both open and closed card sorting methods~\citep{wood2008card}. In the open card sorting phase, researchers create their own groups without predefined categories. We began by interpreting the meanings of different segments within the model names and organizing them into distinct categories based on these interpretations. In the closed card sorting phase, researchers are provided with predefined categories and tasked with sorting data segments accordingly. After categorizing 13\% of 384 model name samples through open card sorting, we applied the resulting categories to the remaining 87\% of the samples using a closed card sorting. Specifically, we iterated through the following steps:

\noindent\textbf{Step 1: Selection of representative samples.} We used a stratified random sampling method with a 95\% confidence level and a 5\% margin error\footnote{https://www.surveymonkey.com/mp/sample-size-calculator/}~\citep{singh2014sampling, cocks2013sample} to select a representative sample of 384 models from a total population of 52,227 for manual analysis. Model names can contain a variable number of segments (see \Cref{subsec:convention_software}), ranging from 2 to N, where N can be any positive integer. We stratified our sampling method according to the number of segments (delimited by forward slashes or hyphens). For instance, a model like mdhugol/indonesia-bert-sentiment-classification\footnote{https://huggingface.co/mdhugol/indonesia-bert-sentiment-classification} has 4 segments, placing it in stratum 4. Each stratum represents a distinct grouping of models that potentially convey different semantic meanings or characteristics based on their naming conventions. Additionally, we ensured that each selected \llm in the sample had a unique owner to prevent overrepresentation from a single source, as we assume that an owner may follow the same naming pattern for all models. Given that a single owner may use the same number of segments in multiple model names, such as AdamCodd/yolos-small-person\footnote{https://huggingface.co/AdamCodd/yolos-small-person}, AdamCodd/donut-receipts-extract\footnote{https://huggingface.co/AdamCodd/donut-receipts-extract}, and AdamCodd/tinybert-sentiment-amazon\footnote{https://huggingface.co/AdamCodd/tinybert-sentiment-amazon}, each having 3 segments, we applied owner uniqueness across all strata.

\noindent\textbf{Step 2: Interpretation and labeling of the model name segments using open card sorting method.} Following the selection of 384 models using a stratified random sampling method, the first and second authors engaged on interpreting the meaning of model names, which are the same as the repository names. To initiate this process, they randomly selected 50 models from the pool of 384 for detailed analysis. This initial subset allows us to gain insight into the diversity and complexity of model naming conventions on \HF, helping us to refine our categorization approach for the larger sample. Each author independently assigned labels to the segments of their assigned model names, drawing upon the semantic information conveyed within the segments. We initially provided an example of model names and their segments in \Cref{inconsistencies}, illustrating five different model names in \Cref{subsec:convention_software},  each containing distinct segments that can be interpreted as the base model, language, variant type, identifier, and version. The first model name in \Cref{inconsistencies}, with its five segments, exemplifies how a single model name can encompass multiple distinct segments. This process highlights the diversity in segment labeling, where different models may have varying numbers of segments, each conveying different types of information.

The terms used in the segments of the model names were clear and self-explanatory, which facilitated the open card sorting process. For instance, terms like ``Llama'' are universally understood to denote a \base, ``7B'' indicates size, and prefixes like ``v\textbackslash d+(\textbackslash.\textbackslash d+)*'' signify version identifiers. Other terms such as ``dataset'' and ``task'' were explicitly mentioned in the model cards. The first two authors familiarized themselves with these labels during an initial observation of 50 repositories, as detailed in \Cref{identify_base}

Following the independent phase, the authors discussed and compared their identified labels. Through a negotiated agreement process~\citep{campbell2013coding}, which involves collaborative discussion and resolution of differences, they addressed all discrepancies and reached a mutual consensus on the most appropriate labels for each model. An example of such discrepancies is when the first author labeled ``20epoch'' as ``epoch,'' and ``direct preference optimization'' as ``features,'' whereas the second author categorized them as ``training mechanisms.'' In this case, after further investigation, they reached an agreement to classify all these keywords under ``training mechanisms.'' This negotiated agreement ensured a consistent labeling scheme for the remaining models at the right level of abstraction, thereby laying the groundwork for robust analysis and interpretation.

\noindent\textbf{Step 3: Closed card sorting for the remaining models.} 
Using the labels identified in the previous step and considering the substantial agreement achieved in Step 2, the first and second authors performed a closed card sorting analysis on the remaining 334 models~\citep{saldana2015coding}. They applied the predefined categories from the open card sorting phase to systematically code the semantic meanings of the segments in the model names. To ensure the reliability of the labeling process, the authors calculated Cohen's Kappa~\citep{vieira2010cohen} to measure inter-rater agreement before reaching a final consensus. Cohen's Kappa evaluates the level of agreement between two raters beyond what would be expected by chance, accounting for random agreement. In this study, Cohen's Kappa achieved a score of 0.74, indicating substantial agreement~\citep{perez2020systematic}. The analysis was conducted using the scikit-learn implementation~\citep{pedregosa2011scikit}. Cohen's Kappa has been widely used for reliability measurement in software engineering research~\citep{perez2020systematic, ali2020quality, yang2021quality}.
 
\emph{Identifying the versioning conventions.} Practitioners on \HF often create new repositories for different model releases instead of evolving versions within the commit history of a single repository. This multi-repo phenomenon implies that version information might be dispersed across multiple repositories rather than being consolidated in one. Thus, our analysis followed such cases by identifying and categorizing version indicators in model names across separate repositories. To identify the model versioning convention, we developed a script \citep{SAILResearch2024} that uses a regex ``v\textbackslash d+(\textbackslash.\textbackslash d+)*'' to extract the version number from all \llm names. We then categorized these versions based on the numerical values following the ``v'' identifier. Versions were grouped as follows: ``major'' for v1, v2, and so on; ``minor'' for v1.x, v2.x where x is greater than 0; and ``patch'' for v1.x.y, v2.x.y where both x and y are greater than 0. It is important to note that the concept of major, minor and patches are not used anywhere on \HF. However, these identifiers resemble those used in software engineering, which is why we followed the same approach to name the identifiers in this manner. Subsequently, we calculated the number of \llms following each practice and analyzed the results.

\emph{Determination of the frequency of file changes on \HF model registry through their commits.} We authored a Python script~\citep{SAILResearch2024} to extract modified repository files within each model repository. This was done to determine implicit model versions. Implicit model versions refer to various alterations made to model weights or configurations, that can only be detected from the commit history of the model's repository or by examining changes in the model's binary files, as they had no accompanying version or model update annotations. After extracting all the files that are associated with \llms on each repository, we focused more on the model binary files used to store parameter information, in particular \textsc{.bin}, \textsc{.pt}, \textsc{.pth}, \textsc{.h5}, and \textsc{.safetensors} (including possible variations). We operated under the assumption that changes to these files signify new versions of the model, potentially resulting in observable differences in model inference behavior. While this assumption is reasonable based on the role of these files in storing model parameters, the actual impact on inference behavior may vary depending on the nature of the changes. As users typically access the latest version of models from \HF, in a similar vein as R users would do when installing R packages~\citep{decan2016github}, this means that they might unknowingly obtain an implicit version that is not documented as a new release and they might not be meant as such. To determine whether a new commit of a model binary file introduces changes, we checked for changes in the file hashes, as these commits do not have accompanying tags like those found on GitHub. Changes in the hashes could result from various factors such as updates to the model architecture, adjustments in hyperparameters, or modifications to the training data.

During our analysis of the files associated with each \llm repository, we encountered a total of 452 unique file extensions. However, numerical extensions such as \textsc{.1}, \textsc{.0}, \textsc{.172}, and others were filtered out to focus on meaningful extensions that could be categorized. It is important to note that none of these numerical extensions were related to model weight files. After this filtering, we have 192 unique file extensions. The primary goal of this categorization was to achieve two main objectives: first, to assess the variety of file types maintained in the \HF model registry, and second, to identify the key file extensions associated with model binaries. By manually categorizing these extensions, we aimed to gain a comprehensive understanding of the file types used, particularly focusing on those relevant to model weights and binaries. This understanding is essential for addressing the research question regarding the \llm versioning practices on \HF.

The first and second authors independently categorized file extensions based on their meanings observed during the initial manual observation of 50 repositories. They agreed on classifying model weight extensions (e.g., .bin, .safetensor) as model files; text extensions (e.g., .md) as documentation; data-interchange extensions (e.g., .json) as configuration; and programming extensions (e.g., .py) as code files. These were the categories identified during the initial observation. This categorization is essential for understanding the convention of files within the \HF \repository, facilitating the efficient management of \llms. In cases of disagreement, such as distinguishing between data and configuration files, the authors reached a consensus through negotiated agreement. 

Following the manual categorization, the authors aggregated the fine-grained file extension label into broader categories. For instance, while initially categorized separately, data and configuration files were combined under ``Data \& Configuration'' since some files serve both purposes, such as JSON files. Ultimately, this aggregation resulted in five distinct categories: ``Code files'' for extensions like \textsc{.py}, \textsc{.cpp}, \textsc{.java}; ``Data \& Configuration files'' for \textsc{.json}, \textsc{.xml}, and similar extensions; ``Documentation files'' for \textsc{.md}, \textsc{.csv}, and others; ``Model binary files'' for \textsc{.safetensor}, \textsc{.bin}, and similar extensions; and ``Other files'' for extensions like \textsc{.jpeg}, \textsc{.zip}. Finally, we visualized the distribution of changes across these file extension categories and present the results in a table.

\emph{Determining the frequency of changes in model binary files.} 
After categorizing all repository files on \HF, we focused on model binary files, as changes in these files could indicate updates or new versions of \llms. We categorized model weight files based on their extensions, such as .safetensor, .bin, and others. Subsequently, we visualized the number of changes occurring in these files. This approach helps us understand the rate at which implicit versions are embedded in \llm repositories on \HF. We created visual representations to illustrate the frequency distribution of changes and computed the mean frequency of changes for each model binary file. This analysis allows us to assess how often model binary files are modified, even in the absence of explicit versioning in the model name.

\subsubsection{Results.}
\begin{figure*}[t]
  \centering
  \includegraphics[width=\textwidth]{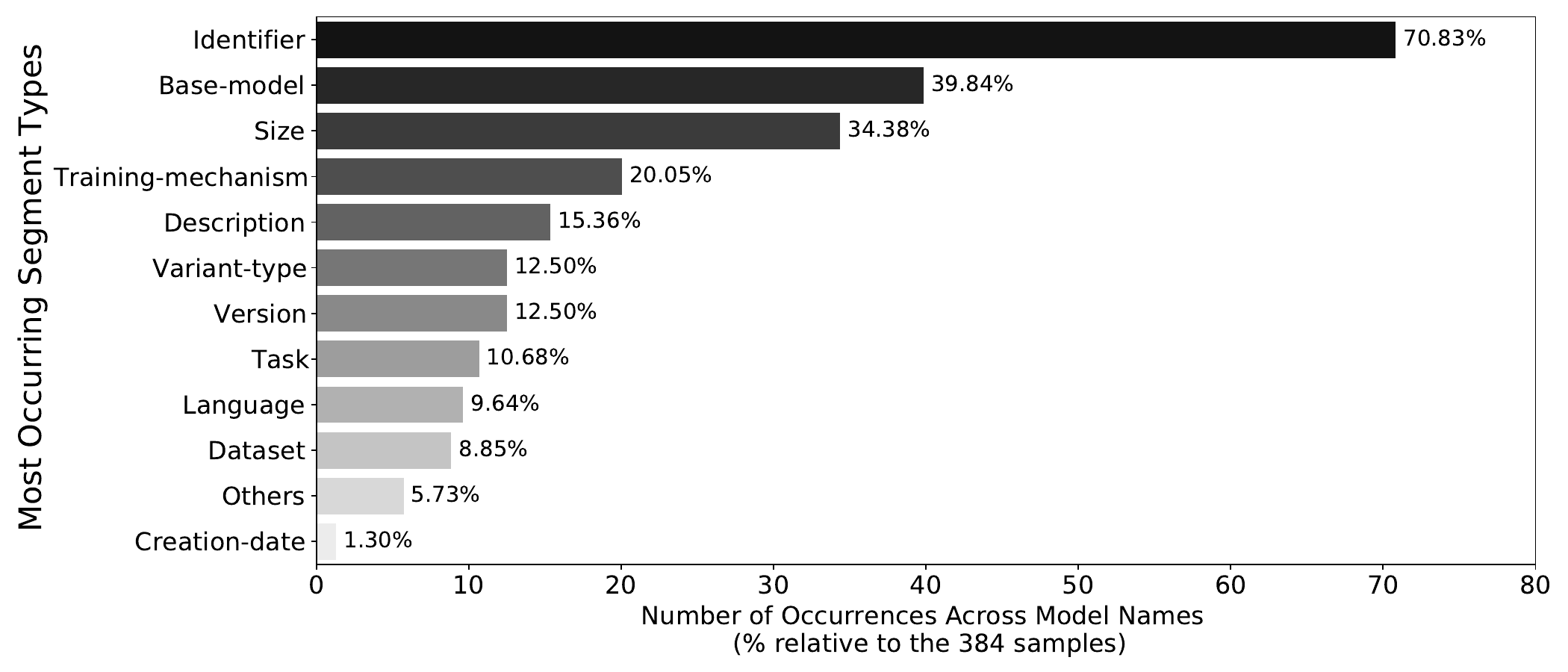} 
  \caption{Visualization of naming convention segment types in 384 manually analyzed model names on \HF. Each model name was broken down into 928 segments, resulting in 12 segment types, which are plotted on the y-axis, with their frequency of occurrences on the x-axis.}
  \label{occuring}
\end{figure*}

\emph{Current naming convention of \llms on \HF.} \textbf{The segments in the naming convention of \llms on \HF encompass 12 segments types, with identifiers (70.8\% of the model names), \base (39.8\%), and size (34.3\%) being the most prevalent.} \Cref{occuring} illustrates the distribution of 12 segment types identified within \HF model naming practices. The y-axis represents the segments types, while the x-axis displays the percentage occurrence of each term relative to the 384 manually analyzed models. We provide a detailed explanation of each segment type in our study.

\begin{itemize}
    \item \textbf{Identifier}: An identifier, such as ``myllm'' in ``truemansquad/myllm,"\footnote{https://huggingface.co/truemansquad/myllm} aims to uniquely differentiate a model, similar to how variable and method names try to distinguish variables and methods based on their semantics. Since an identifier by itself might not be unique across the overall \HF model registry, it may also be combined with additional contextual information to further specify and identify models.
    \item \textbf{Base model}:``bart'' in ``voidful/bart-distractor-generation-both''\footnote{https://huggingface.co/voidful/voidful/bart-distractor-generation-both} is an example of a \base. It represents the \base from which the custom model was adapted, serving as the foundational architecture or framework from which adaptations are made to suit specific tasks or applications. 
   \item \textbf{Size}: In ``Voicelab/trurl-2-13b,"\footnote{https://huggingface.co/Voicelab/trurl-2-13b} the designation ``13b" signifies the number of parameters, indicating the computational capacity of the underlying \llms. 
    \item \textbf{Description}: The inclusion of ``my\_awesome\_qa\_model" in ``lakshyasoni/my\_awesome\_qa\_model"\footnote{https://huggingface.co/lakshyasoni/my\_awesome\_qa\_model} provides a description chosen by the model's owner, emphasizing the model's purpose as a question answering model. Unlike identifiers, which are always single words, descriptions are typically meaningful sentences or phrases that convey more detailed information about the model.
    \item \textbf{Variant type}: ``finetune" in ``gsomers-smarsh/distilgpt2-emailtype-finetune"{\footnote{https://huggingface.co/distilgpt2-emailtype-finetune}} exemplifies an adaptation method applied to the \base. This indicates a specific adaptation, such as fine-tuning for a particular task or domain.
    \item \textbf{Version}: The presence of ``v1" and ``v1.2" in ``Haary/TinyLlama-1.1B-usk-v1"\footnote{https://huggingface.co/Haary/TinyLlama-1.1B-usk-v1} and ``SQAI/distilroberta-base\_finetune\_v1.2"\footnote{https://huggingface.co/SQAI/distilroberta-base\_finetune\_v1.2} denote different versions of the models. They signify distinct iterations of the \dmodels, reflecting updates, improvements, or changes made over time to address user feedback or evolving requirements.
    \item \textbf{Training mechanism}: The ``loss\_5e-06" in ``Shijia/furina\_pan\_loss\_5e-06"\footnote{https://huggingface.co/Shijia/furina\_pan\_loss\_5e-06} is an example of a training mechanism. It specifies the training hyperparameter or methodologies applied during the model training process, such as the choice of loss function or optimizer settings, influencing model performance and convergence. Under this category, we also classified all the information regarding training task, tuning method, optimization technique, programming, scaling, instruct, direct preference optimization (dpo), post-training quantization for generative pre-trained transformer (GPTQ), Python, and upscalled. ``Instruct" refers to a training mechanism that involves providing specific instructions or directives to the model during the training process, guiding its learning behavior. ``DPO" is a training mechanism aimed at optimizing the model's parameters directly based on user preferences or desired outcomes, bypassing intermediate steps or metrics. ``GPTQ" is a technique used to quantize or compress the parameters of a pre-trained transformer model after the training phase, reducing its memory footprint or computational requirements while preserving performance. ``Python" indicates that the model was trained using the Python programming language, commonly used for developing ML models and frameworks. ``Upscaled" denotes a training mechanism where the model's capacity or size is increased, often resulting in improved performance or capabilities, such as increased resolution or feature representation.
    \item \textbf{Task}: In ``youdiniplays/filipinolingo\_translation,"\footnote{https://huggingface.co/youdiniplays/filipinolingo\_translation} ``translation" denotes the task for which the model is modified for. It clarifies the model's intended use case or functionality, guiding users in selecting appropriate models for specific tasks or applications, such as language translation.
    \item \textbf{Dataset}: ``indonlu" in ``andikamandalaa/indobert-base-uncased-finetuned-indonlu-smsa"\footnote{https://huggingface.co/andikamandalaa/indobert-base-uncased-finetuned-indonlu-smsa} represents the dataset used for training the model. It provides transparency regarding the training data sources, enabling users to assess the model's domain relevance and generalization capabilities.
    \item \textbf{Language}: In ``abiatarfestus/marian-finetuned-en\_ng\_bible-en-to-ng,"\footnote{https://huggingface.co/abiatarfestus/marian-finetuned-en\_ng\_bible-en-to-ng} ``en" and ``ng" represent the languages for which the model is designed or supports. It ensures compatibility with language-specific tasks or datasets, facilitating seamless integration into language-centric applications.
    \item \textbf{Others}: In ``Arkong/chatglm2-6b-torchkeras-2epoch-11-15,"\footnote{https://huggingface.co/Arkong/chatglm2-6b-torchkeras-2epoch-11-15} the segment ``11-15" illustrates ambiguity within the naming practice. This lack of clarity may hinder users' understanding of the model's attributes or specifications, highlighting the importance of clear and precise terminology.
    \item \textbf{Creation-date}: The inclusion of ``2024-03-01" in ``thrunlab/Mistral\_Sparse\_refined\_web\_relu\_2024-03-01"\footnote{https://huggingface.co/thrunlab/Mistral\_Sparse\_refined\_web\_relu\_2024-03-01} indicates the model's creation date or any other information associated with a date. It provides temporal context and facilitates version control, enabling users to track model evolution and updates over time.
\end{itemize}

Furthermore, the  key segment types indicated in the naming practice contribute to the diversity and inconsistency within the \HF repository. This inconsistency poses challenges for users attempting to quickly and accurately understand model characteristics, which is essential not only for selecting but also for effectively utilizing the most suitable models for their specific tasks. Interestingly, our analysis shows that the `version' segment, which closely resembles semantic versioning, is utilized in only about 5\% of the cases depicted in \Cref{occuring}, highlighting a significant finding. To address these challenges, a more structured and enforceable representation, akin to semantic versioning, could be implemented through model cards or configuration files, providing users with clearer and more standardized information. If this information is already included in model cards or configuration files, it should be consistently indicated there rather than relying on the model names alone.

\textbf{The average download rates of PTLMs mentioning only model size in their name is more than 11 times as high as those mentioning only base model.} Given the prevalence of base model and size in the studied model names, we investigate their association with the number of model downloads. To do this, we compute the download rates of models that include either size, base model, or both in their names, and compare them with the download rates of models that excluded these segments. Therefore, our analysis shows that the download rate for PTLMs that mentioned base models averaged 101.88 downloads, while those that mentioned size averaged 1,163.21 downloads per model. In contrast, models that mentioned both size and base model averaged 314.2 downloads per model, whereas those that mentioned neither averaged only 67.1 downloads per model. This observation suggests a correlation between including base model and/or size in model names and higher download rates than those without these segments, potentially reflecting users' demand for clarity in model names to better assess the relevance or impact of version updates.

\begin{figure*}[t]
\centering
\includegraphics[width=\textwidth]{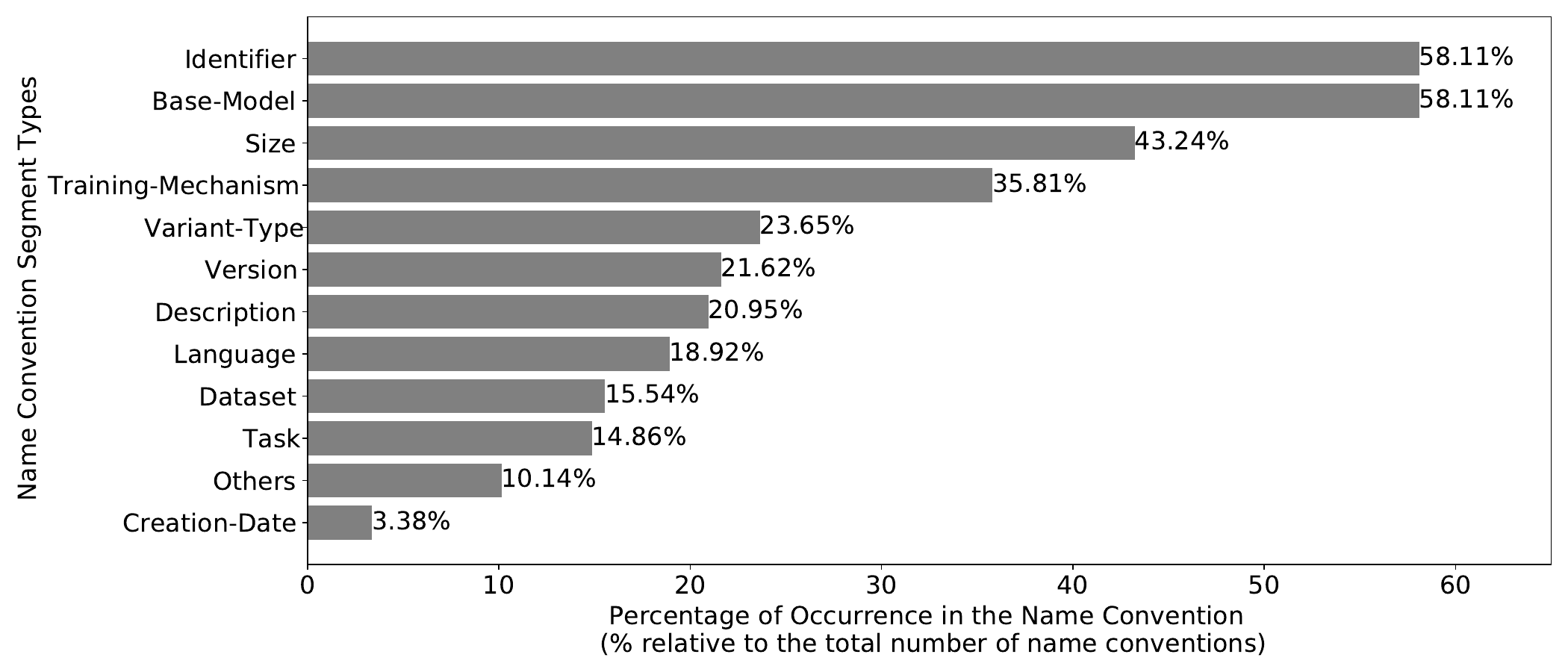}
\caption{Visualization of labeled segments from 384 manually analyzed model names on \HF. Each model name is composed of various segment types, as illustrated by the naming convention \{base-model\}\{variant-type\}\{dataset\}. The dataset, consisting of several segments in the 148 naming conventions, is plotted with the segment types on the y-axis and the number of times each segment type appears in these conventions on the x-axis.}
\label{element}
\end{figure*}

\textbf{Our manual analysis reveals the existence of 148 distinct name convention for repositories on \HF,} highlighting a notable level of  diversity in model naming strategies across \HF model's repositories.
\Cref{naming_convention} presents the 20 most prevalent name convention, including the percentage of repositories utilizing each convention, along with illustrative examples. Despite the multitude of conventions, the predominant formats observed are \{identifier\}, \{identifier\}\{size\}, and \{\base\}\{identifier\}. 

This diversity in naming practices may reflect a range of practitioner preferences and practices. Names are not enforced on \HF or other registries, nor are there official standards, leading to the observed inconsistencies. The discrepancy between what practitioners prefer or would ideally like when it comes to naming PTMs and the actual naming convention that are commonly used in practice highlights the need for more standardized guidelines to ensure consistency and clarity in PTM naming~\citep{jiang2024naming}.

\begin{table}[t]
    \centering
    \caption{The 20 (out of 148) most occurring naming convention and examples, comprising at least 3 models}
    \label{naming_convention}
    \begin{tabular}{p{7cm}rl}
        \toprule
        \textbf{Naming convention} & \textbf{\% Model} & \textbf{Examples} \\
        \midrule
        \{identifier\} & 17.96 & vesteinn/ScandiBERT \\
        \{identifier\}\{size\} & 8.07 & automerger/Experiment27-7B \\
        \{\base\}\{identifier\} & 3.12 & eugenesiow/bart-paraphrase \\ 
        \{description\} & 2.86 & FrankTCH/Trans-from-scratch \\
        \{identifier\}\{version\} & 2.6 & anupk/AskPaul-V2 \\
        \{identifier\}\{description\} & 2.34 & kamel-usp/aes\_enem\_models-sourceA-ordinal- \\
        & & from-bertimbau-large-C1 \\
        \{identifier\}\{\base\}\{size\} & 2.08 & cerebras/Cerebras-GPT-111M \\
        \{task\} & 2.08 & bgoel4132/tweet-disaster-classifier \\
        \{\base\}\{training-mechanism\} & 1.82 & danielkty22/gpt2-ep-1.4-b-4-lr-4e-06- \\
        & & dp-0.1-ss-0-st-False-fh-False-hs-200\_normal \\
        \{identifier\}\{\base\} & 1.56 & nbroad/ESG-BERT \\ 
        \{identifier\}\{task\}	& 1.56	& emarron/JARVIS-email-sorter\\
        \{identifier\}\{size\}\{training-mechanism\}	& 1.30	& Ichigo2899/Airoboros-13b-8k-TGI-GPTQ\\
        \{identifier\}\{training-mechanism\}	& 1.30	& ATYOSHIDA/Q2\_default\_7030\_seed42\_random\_2\\
        \{base-model\}\{variant-type\}\{dataset\}	& 1.30	& Quocc/roberta-finetuned-subjqa-movies\_2\\
        \{base-model\}\{size\}	& 1.04	& openerotica/Qwen-7b\\
        \{identifier\}\{others\}	& 1.04	& controltensor/subnet-model-19\\
        \{identifier\}\{base-model\}\{training-mechanism\}	& 1.04	& danielkty22/probe\_gpt2-medium-ep-1.0-b-4-lr-1e\\
        & & -05-dp-0.01-ss-0-st-False-fh-False-hs-100\\
        \{identifier\}\{size\}\{description\}	& 1.04	& vincentmin/bloomz-1b1-eli5-reward\\
        \{base-model\}\{size\}\{identifier\}	& 1.04	& win10/Qwen1.5-0.5b-Xia-Ai\\
        \{identifier\}\{base-model\}\{dataset\}	& 0.78	& mesolitica/malaysian-mistral-mmmmodal\\
        
        \bottomrule
    \end{tabular}
\end{table}

\textbf{Our analysis reveals that identifiers, base models, and size are the most prevalent naming segments in the naming conventions used by practitioners on Hugging Face.} \Cref{element} shows the prevalence of each segment type  in the observed naming convention patterns. Even though identifiers can be arbitrary and are at the practitioners’ discretion, the most prevalent segments among the practitioners aside from ``identifiers” are: ``base model", ``size", ``training mechanism", ``variant type", and ``version". This indicates that practitioners prioritize communicating key model characteristics such as architectural foundation, scale, training methodology, and specific variant details to provide quick, comprehensive insights into the model’s essential attributes and differentiation.

This result corroborates the findings in \citep{jiang2023exploring}, which revealed a preference of practitioners for the architecture, size, and task naming segments. Building on their observations, we found that, due to the unique characteristics of language models compared to other model types, practitioners on Hugging Face also emphasize training mechanisms and version. These segments, which were shown to have low prevalence in Jiang et al.’s study, appear to play a more prominent role in the naming practices for language models. This may be one of the aspects that distinguish language models from other types of models.

Note that, what \citep{jiang2023exploring} referred to as naming conventions in their study is categorized as naming segments in our work, as 148 naming conventions in our paper comprise one or more segments types. Additionally, while \citep{jiang2023exploring} studied models of different domains, our focus was specifically on language models. The insights from our analysis highlight the need for further studies focusing on naming conventions specific to models of different domains.

\textbf{Our analysis of naming conventions reveals that \{identifier\}\{base-model\}\{size\}, \{identifier\}\{size\}, and \{identifier\}\{training-mechanism\} are associated with the highest download rates among the prevalent conventions. However, no significant relationship was found between the length of model names and download rates.}
We analyzed the relationship between download rates and naming conventions. To achieve this, we first ranked the naming conventions by their average download rates. The naming conventions—such as \{identifier\}\{size\}\{language\}\{version\}, \{identifier\}\{size\}\{language\}, \{base-model\}\{size\}\{variant-type\} \{description\}, \{identifier\}\{base-model\}\{version\}, and \{base-model\}\{size\}\{dataset\}\{training-mechanism\}—had the highest average download rates (19,434, 5,762, 3,015, 2,522, and 2,418 downloads per model, respectively). However, these naming conventions lacked sufficient data points to draw a definitive conclusion. Consequently, we filtered out naming conventions that were used by fewer than three PTLMs, resulting in 20 naming conventions (13.5\% of 148) being qualified for this analysis.

\begin{figure*}[t]
\centering
\includegraphics[width=0.7\textwidth]{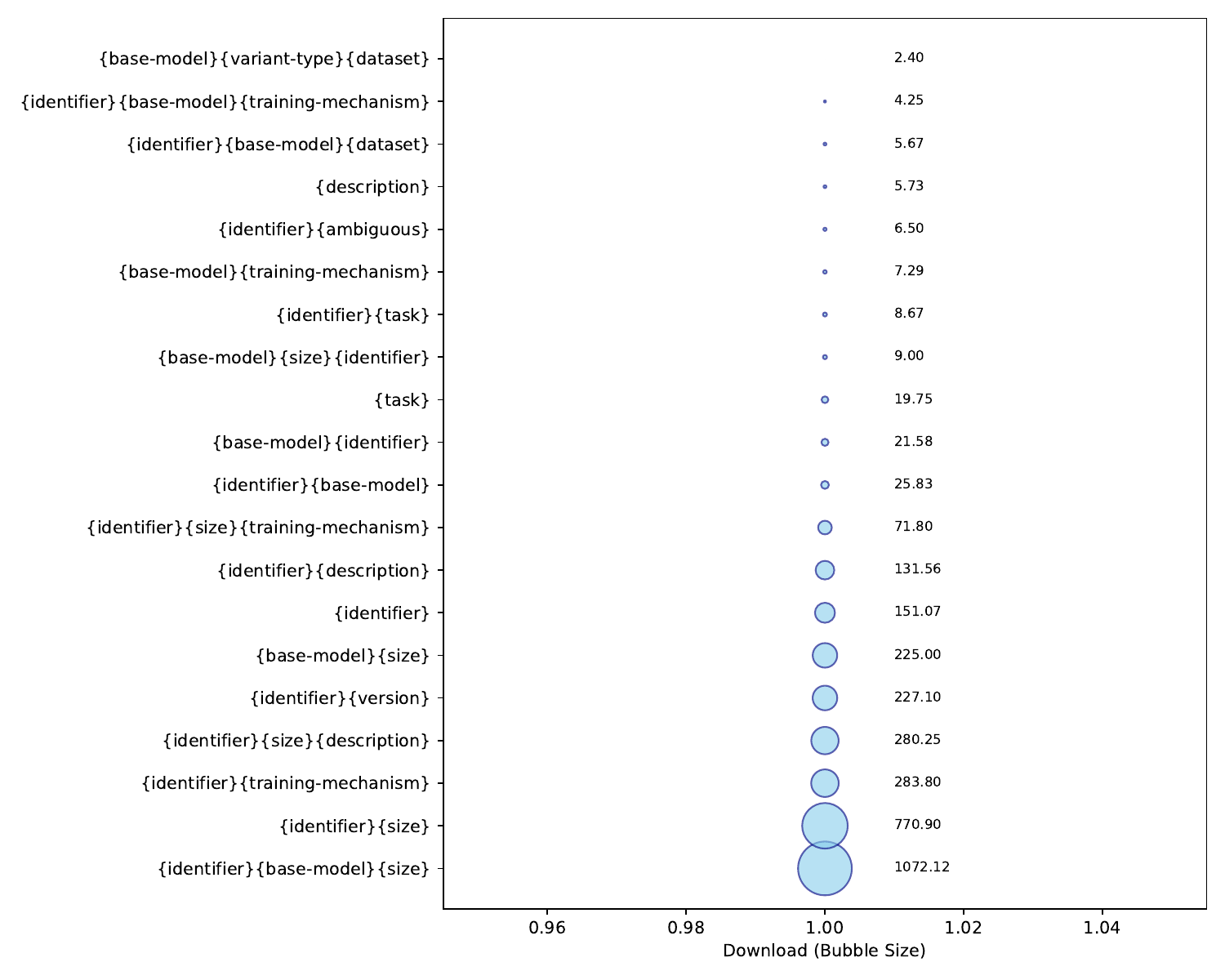}
\caption{Visualization of the top 20 prevalent naming conventions out of 148 and their average download rate per model for each naming convention. The naming conventions are displayed on the y-axis, while the normalized download rate is shown on the x-axis. The higher the download rate, the larger the bubble.}
\label{convention_download}
\end{figure*}

\Cref{convention_download} visualizes the download rates of the prevalent naming conventions using a bubble plot. The y-axis represents naming conventions, while the x-axis indicates the normalized size of each bubble, and the bubble size reflects the average download rate. The results show that \{identifier\}\{base-model\}\{size\}, which averages 1,072.12 downloads per model; \{identifier\}\{size\}, averaging 770.90 downloads; and \{identifier\}\{training-mechanism\}, averaging 283.80 downloads, are the most widely adopted naming conventions with notable download rates. This analysis highlights the frequent inclusion of base model, size, and training mechanism segments in naming conventions with high download rates, suggesting that these features are key factors in naming preferences. These findings align with the results in \citep{jiang2023exploring}’s study, which revealed participants’ preference for naming models are based on architectural characteristics and intended functions.

To quantify the relationship between model name length and download rates, we conducted a Pearson correlation analysis and a linear regression. The Pearson correlation coefficient (-0.024) and p-value (0.7718) indicate no statistically significant correlation between the two variables. The negative sign of the correlation coefficient suggests a negligible tendency for download rates to decrease as model names lengthen. The linear regression analysis further supports this observation, with a slope of -0.024 indicating a slight negative relationship between name length and download rates. However, this effect is minimal, and the near-zero intercept reflects the minimal correlation of name length with download rates, highlighting that even very short names do not significantly alter the former. The R-squared value of 0.0006 shows that only 0.06\% of the variance in download rates can be explained by model name length, emphasizing that factors other than name length, such as number of direct descendant model and documentation quality \citep{jones2024we}, play a more significant role in influencing download rate as reported in Jones et al. study.

\emph{Current Versioning Conventions on HF.}
\textbf{Only 3,471 (6.64\%) of 52,227 \llms have a version segment in the name convention}. Our analysis of model name conventions on \HF reveals that the percentage of \llms specifying the version of the model is very low. This lack of versioning information in model names highlights the potential difficulties in tracking and managing different iterations of \llms on the platform. Adding to the complexity, each version of a model on \HF is often placed in a separate repository, further underscoring the need for proper naming conventions and versioning tags to help users easily identify and differentiate between various iterations. This low percentage might suggest that practitioners only specify versions when necessary. It is possible that many practitioners are adapting the base model to publish a new \llm once without further corrections or updates, indicating a one-off customization for specific use cases rather than ongoing iterative development. Alternatively, the lack of versioning might reflect a broader issue where there are no established guidelines or enforced standards for naming conventions on \HF. This lack of enforcement and standardization could contribute to the overall low rate of version specification in model names.

\noindent\textbf{Among the \llms that do include version information in their name convention on \HF, the predominant strategy is major versioning, following the software development standard of identifiers like v1, v2, etc. Specifically, we observed that 67\% of these \llms adopted this major versioning approach. This indicates that when version information is provided, it predominantly reflects major versions only.} Our analysis reveals that major versions (e.g., v1, v2) are the predominant versioning practice on \HF among \llms that specify versions, accounting for 67\% of the 3,471 \llms with version information in their names. While major versioning is widely adopted for updating many models, the limited use of minor versions (e.g., v1.1) for 33\% models suggests deviations from established software engineering practices. Minor versions provide finer granularity in version updates, facilitating more precise tracking of incremental changes and the ability to check compatibility with downstream applications \citep{paez2018versioning}. Enhancing the use of minor versions on \HF could align practices more closely with industry standards, which emphasize the importance of both major and minor version distinctions. \\

\takeaway{Summary}{\HF \llms feature diverse naming practices (148) on \HF, composed of segments matching 12 possible segments with segments representing ``identifiers,'' ``\base,'' and ``size'' the most frequent indicated in the names. \textit{Major} versioning identifiers (67\% of 3,471 models) dominates.}

\emph{Categories of Changed Files in Repository Commits.} \textbf{A total of 1,282,874 changes were observed across 52,227 \llm repositories. However, only 3,471 of these changes are explicitly communicated through version identifiers in the model names, leaving other significant changes implicit within the model's repository.} Furthermore, we observed that many practitioners prefer using separate HF repositories for different major and minor releases instead of evolving versions within the commit history of a single repository. This practice further complicates tracking and managing different iterations of \llms. \Cref{change_frequency} shows the categories of files that are changed on \HF, the frequency of changes in percentage, the number of \llms (in percentage) that made changes to each file category, and the average number of changes per model. The mean changes per model are determined by dividing the change frequency by the total number of models that made the changes. It is evident that different file categories exhibit distinct conventions of change.

Model files, referring to model binary files, have the highest frequency of changes (40.87\%), with a substantial proportion of models making changes to these files (32.81\%). Despite this high frequency, only 0.66\% of these changes are reflected in the model names through version identifiers. This implies that approximately 40.21\% of changes are implicit, not declared in the model names or anywhere in the repository. This high percentage of implicit versions suggests a significant oversight in versioning practices, similar to issues seen in R where users might face difficulties due to inadequate versioning of packages\citep{decan2016github}. The frequent updates to model binaries, likely driven by ongoing improvements and optimizations, highlight the need for more structured versioning that includes major, minor, and patch revisions. The high average of 12 changes per model in this category suggests that adopting a more rigorous versioning approach could help mitigate confusion and compatibility issues for users.

Data \& Configuration Files also exhibit a high frequency of changes (34\%) and are similarly important, reflected in the comparable percentage of models making changes to these files (32.85\%). This suggests that modifications in configuration and data are frequent and essential for maintaining and enhancing model performance. The average changes per model in this category (10) are significant, highlighting the ongoing need to update and refine configuration settings and data inputs.

Other Files, despite having a moderate frequency of changes (14\%), see a relatively low percentage of models making changes (6.95\%). This discrepancy could imply that when changes do occur in these files, they are often more substantial or involve fewer models but with more significant changes per instance. The high average changes in this category (19 changes per model) support this notion, indicating substantial modifications when changes are made.

Documentation files, while not changed as frequently (11\%), are updated in a considerable proportion of models (25.81\%), highlighting the importance of maintaining accurate and up-to-date documentation. The lower average changes (4 per model) suggest that documentation updates are more straightforward and less frequent compared to other file categories. The reasons for these updates could include initial documentation being incomplete, unclear, or missing, among other possibilities.

Lastly, Code Files have the lowest frequency of changes (1\%) and a minimal percentage of models making changes (1.58\%). This is not surprising, given that \HF primarily functions as a model registry for downstream tasks rather than a code repository like GitHub. Consequently, there is less need for frequent updates to code files. The low average changes (4 per model) suggest that code updates are infrequent and involve smaller adjustments rather than large-scale revisions.

\begin{table}[t]
    \centering
    \begin{tabular}{lrrr}
        \hline
        \textbf{File Categories} & \textbf{CF (\%)} & \textbf{NM (\%)} & \textbf{ACPM} \\
        \hline
        Model Files & 40.878 & 32.81 & 12 \\
        Data \& Configuration Files & 33.946 & 32.85 & 10 \\
        Other Files & 13.825 & 6.95 & 19 \\
        Documentation Files & 10.658 & 25.81 & 4 \\
        Code Files & 0.69 & 1.58 & 4 \\
        \hline
    \end{tabular}
    \caption{Change Frequency and Average Number of Changes per Model by File Category. CF(\%): Percentage of Change Frequency, NM (\%): Percentage of Number of Models (relative to 52,227), ACPM: Average Number of Changes Per Model.}
    \label{change_frequency}
\end{table}

\emph{Categories of Changed Model Binary Files in Repository Commits.}
\textbf{Frequent changes (a total of 524,419 changes) are observed in the model weight files of 52,227 models, and security-focused tensor files are the most commonly used ML framework for storing model weight on \HF, exhibiting the highest frequency of changes, averaging 7.88 changes per model.}

We identified various model binary file extensions used for storing model weights on \HF, associated with different ML frameworks: Generic Binary files (.bin, .model, .mdl), PyTorch model files (.pt, .pth, .torch), TensorFlow model files (.meta, .ckpt, .pb), TensorFlow Lite model files (.tflite), ONNX model files (.onnx), Apple Core ML model files (.mlmodel), and Security-focused tensor files (.safetensors). \Cref{mofel_file} presents these model file categories along with their frequency of changes, percentage of models utilizing each category, and average changes per model.

Security-focused tensor files exhibit the highest frequency of changes, averaging 7.88 changes per model, indicating significant maintenance and updates. Generic Binary files show substantial activity with an average of 4.18 changes per model, widely utilized across models. PyTorch model files demonstrate a notably high average of 31.92 changes per model, reflecting dynamic development despite lower utilization. ONNX model files average 4.28 changes per model, emphasizing their role in interoperability. TensorFlow and TensorFlow Lite model files exhibit lower activity with averages of 3.03 and 3.53 changes per model, respectively. Apple Core ML model files have the lowest frequency of changes at 1.44 per model, reflecting their specialized use within the Apple ecosystem.\\

\begin{table}[t]
\centering
\caption{Categorization of model weight files. Percentage of Change Frequency, NM (\%): Percentage of Number of Models (relative to 52,227), ACPM: Average Number of Changes Per Model.}
\label{mofel_file}
\begin{tabular}{lrrr}
\toprule
\textbf{Model File Categories} & \textbf{CF (\%)} & \textbf{NM (\%)} & \textbf{ACPM} \\
\midrule
Security-focused tensor file & 62.63 & 53.63\% & 7.88 \\
Generic Binary file & 27.17 & 43.85\% & 4.18 \\
PyTorch model file & 9.96 & 2.10\% & 31.92 \\
ONNX model file & 0.20 & 0.32\% & 4.28 \\
TensorFlow model files & 0.02 & 0.05\% & 3.03 \\
TensorFlow Lite model file & 0.01 & 0.02\% & 3.53 \\
Apple Core ML model file & 0.01 & 0.03\% & 1.44 \\
\bottomrule
\end{tabular}
\end{table} 

\takeaway{Summary}{We highlight a significant disconnect between versioning practices and model release activities (3,471 declared versions instead of potentially 524,419 versions, if each change is considered a potential version). Frequent changes (an average of 12 changes per model) were observed in model files more than in configuration, documentation, and code files, but these modifications weren't always reflected in model names or version identifiers (3,471 out of 524,417 were reflected). Similarly, up to seven different ML frameworks are utilized for storing model weights on \HF, indicating potential implications for model interoperability and user-friendliness concerning versioning conventions. Security-focused tensor files exhibit the highest frequency of changes among all categories, suggesting intensive maintenance and updates \citep{singla2023machine}. Code and documentation file categories have the fewest changes (4 per model), while model files have the most changes (12 per model), underscoring frequent modifications that can significantly impact model versions.}

\subsection{\textbf{RQ-2:} What are the \vtypes and their qualities on HF} 

\subsubsection{Motivation} 
This research question investigates the reproducibility and transparency of \llm releases on \HF. Understanding these aspects is important because they affect users' trust in the consistency and dependability of the models. Transparency is defined in terms of the availability of model cards and dataset documentation, while reproducibility pertains to understanding the provenance of \llms and their variant types on \HF. For transparency, we explore the rate at which practitioners release model cards with their \llms and how frequently they mention the datasets used to train their \llms. For reproducibility, we investigate the number of base models adapted to publish the 52,227 \llms we are studying and how often practitioners state the adaptation methods, which result in different variant types. This understanding is supported by the availability of configuration files that specify model settings and base model details. Inconsistent release practices may hinder users and developers not only in selecting models for their downstream applications but also in assessing model reliability and efficacy before deployment. By studying transparency and reproducibility for different model variants, we aim to identify areas where improvements are needed to enhance the overall model sharing and reuse process on \HF. This information is valuable for both model creators and users, ensuring that models are more accessible and easier to integrate into various applications.

\subsubsection{Approach} 
\emph{\textbf{1.) Provenance of \llms on \HF (reproducibility).}}
To investigate the reproducibility of each \llm release, we first explored their provenance of \llms by identifying the \base within the configuration file of each \llm release in its respective repository. To extract configuration information of a release, we leveraged the \texttt{config.values()} function of the \HF Transformers library. At the time of collecting the data, not all models have configuration files, such as \texttt{nvidia/retro-8b-base-4k}, indicating the owner did not upload them using either the standard \textit{HfAPI}\footnote{https://huggingface.co/docs/huggingface\_hub/en/package\_reference/hf\_api} or the \textit{transformers.PretrainedConfig}\footnote{https://huggingface.co/docs/transformers/en/main\_classes/configuration} class. Standard \textit{HfAPI} is a part of the \HF Hub that provides a unified interface for accessing model configuration information, while \textit{transformers.PretrainedConfig} is a class within the \HF transformers library specifically designed for handling model configurations. We calculated the percentage of \llm models that have configuration files in their repositories.

\noindent \emph{\textbf{2.) Variant types of \llms on \HF (reproducibility).}} We identified the variant types from the model name. As explained in \Cref{subsec:pre-trained-llms}, \textit{Variant types} encompass types of modification methods applied to the \base, such as Fine-tuning to derive a variant model. We discovered in RQ$_1$ that some keywords, such as ``finetuned", are stated in the names of some models on \HF, and can be classified as the variant type. Therefore, to comprehensively identify all these keywords, we used both manual and automatic methods to analyze the model names, config files, and metadata associated with each of the models in the \HF \repository. We explored only these aspects because both the first author and second author independently examined 50 randomly selected \llms repositories and found that these aspects at least consistently provide the necessary information to identify variant types accurately across those models through the common naming practices. We acknowledge that model cards may sometimes contain the necessary keywords; however, not all models have model cards, and accessing them for some repositories requires manual authentication and waiting time, reducing our sample size. For example, accessing meta-llama/Llama-2-7b-chat\footnote{https://huggingface.co/meta-llama/Llama-2-7b-chat} and meta-llama/Llama-2-7b\footnote{https://huggingface.co/meta-llama/Llama-2-7b} requires manual authentication and waiting time to be granted access by the owner to these \llms.

For our manual analysis, as it is impractical to analyze 52,227 models, we focused on a statistically significant sample of 384 models (confidence level of 95\% percent and margin of error 5\%). For the selected samples, we manually explore different aspects of the models by browsing each model's repository one by one to identify an indications of variant types. First, we focused on the model names, as many models have indications, such as ``finetuned," directly in their name segments. Second, we explored the configuration files to determine whether variant-related segments are present. However, we found that 2\% of the studied \llms lacked configuration files, such as \textit{Sosaka/Alpaca-native-4bit-ggml} and \textit{Skaczmarj/resnet50-truncated.tv\_in1k}, making it impossible to locate this information from them. Third, we explored the tags of each model repository. While some models, like \textit{01-ai/Yi-34B-Chat-4bits}, specified a keyword (\textit{4bits}) in their tags, others, like \textit{upstage/SOLAR-0-70b-8bit}, did not. Subsequent to the manual analysis, we automatically collected these segments from the segments of the model names using a Python script \citep{SAILResearch2024}.

To determine the prevalence of model reproducibility in terms of \vtypes, we calculated the distribution of \llms across \vtypes. This reproducibility characteristic allows us to gain insights into which \vtypes are more prevalent on \HF, thereby helping us understand the common practices in specifying model variants.

\noindent\emph{\textbf{3.) \llms training dataset indication (transparency).}}
To gain a comprehensive understanding of release transparency on HF, we examined how frequently the sources of training datasets are mentioned in the release documentation or in the dedicated areas of the \llm repositories on \HF. This focus is important because the training dataset can significantly impact a model's performance and suitability for specific tasks. Our assessment of the distribution of \llm models that mention training dataset sources involved a dual methodology: manual and automatic.

\noindent\emph{Automatic Method for Identifying \llm Releases that Mentioned the Training Dataset and their sources.} We leveraged the \textit{cardData.datasets} function of the \textit{HfAPI} to extract dataset information from all analyzed releases. Our script retrieves the dataset specified by the model owner. For instance, executing our script on the \llm named ``rhaymison/cuscuz-7b"\footnote{https://huggingface.co/rhaymison/cuscuz-7b} returns ``rhaymison/questions\_answers\_geo\_nord", indicating the dataset used for training the model.

\noindent\emph{Manual Method for Identifying \llm Releases that Mentioned the Training Dataset and their Sources.} We conducted a manual analysis of repositories where the automated technique failed to identify a specified dataset. To ensure an unbiased approach, we considered all \llm releases (25,807 in total) where the automated method did not retrieve any dataset. From this pool, we randomly sampled 379 unique \llms for further analysis using a Python script, with a confidence level of 95\% and a \% margin of error. The selection criteria required each sample to have a unique owner, a unique model, and an available model card for each release. These criteria were collaboratively established by both authors to ensure consistency.

Furthermore, dataset names can be duplicated, while only the source clarifies the exact data used. For example, the model card for saicharan8/telugu-summarization-umt5-small\footnote{https://huggingface.co/saicharan8/telugu-summarization-umt5-small} stated, ``This repository is a fine-tuned version of google/umt5-small\footnote{https://huggingface.co/google/umt5-small} on a Telugu-News Article Summaries Dataset," indicating the utilization of the Telugu-News Article Summaries Dataset for training without mentioning the source. If the source is unavailable, the user cannot access the training dataset, particularly if it is not on \HF. In another example, the model card for Salesforce/codegen2-16B\footnote{https://huggingface.co/Salesforce/codegen2-16B\_P} mentioned, ``This checkpoint is trained on the stricter permissive subset of the deduplicated version of the Stack dataset (v1.1)," accompanied by a link\footnote{https://huggingface.co/datasets/bigcode/the-stack-dedup} to the dataset. Some repositories, such as ``chihoonlee10/T3Q-Merge-SOLAR12"\footnote{https://huggingface.co/chihoonlee10/T3Q-Merge-SOLAR12}, lack any information regarding the training dataset or source. Subsequently, we categorized the \model based on whether they specified the dataset without source, specified the dataset with source, or didn't specify either.

We highlight that our initial sample selection encountered inaccessibility issues with three repositories ``(venkycs/ZySec-2B-v2", ``deepnetguy/gemma-54", and ``deepnetguy/gemma-55") potentially deleted or renamed on \HF. To ensure data integrity, we re-examined all repositories using a Python library, assigning a ``success" status to accessible ones and an ``error 404'' status to unavailable ones. This process identified 95 inaccessible repositories (0.2\% of the total), indicating that model registry releases can be relatively brittle. The fact that between the start and end of our study, 95 out of 52,227 repositories were no longer accessible, could indicate potential disruption of user workflows and applications. We then proceeded with the analysis using the final, accessible sample.

Following this procedure, both the first and second authors independently performed the entire manual analysis by directly accessing each repository to read the model card for the purpose of identifying the mentioned training dataset information. The inter-rater agreement between them was measured using Cohen's kappa score, which was found to be 0.98. Notably, there was only one discrepancy: in a case where a model card stated, ``This model is a fine-tuned version of bert-base-uncased on an unknown dataset," the first author accidentally interpreted the dataset name as ``Unknown," while the second author interpreted it differently. 

\noindent\emph{\textbf{4.) \llms model card publication (transparency)}} 
To investigate the transparency of \llm releases on \HF regarding model card availability, we developed a Python script using the \HF API. This script checks each \llm to determine if it has an associated model card. If a model card is found, the script retrieves it; otherwise, it outputs a message stating, ``Repo card metadata block was not found. Setting CardData to empty." Following this, we analyzed the distribution of \llms, categorizing them based on whether they have model cards or not.

\subsubsection{Result}
\paragraph*{Reproducubility of \HF releases based on provenance.}

\noindent\textbf{98\% of the studied \llms have configuration files that detail the base models adapted for \llms.} Our analysis shows that practitioners make it easier for users to locate the parent model of their current model by including configuration files that detail model origin and other information, such as parameter settings and transformer versions. Furthermore, the variant type of the source model is also documented in these files, though not in all cases. While 98\% have configuration files, only 0.085\%, such as \textit{mlx-community/Mistral-7B-Instruct-v0.2-4-bit} and \textit{01-ai/Yi-34B-Chat-4bits}, actually contain indications of variant type in these files. However, repository names remain important because they provide a quick reference to models and help users differentiate between them, especially when configuration files do not consistently contain all the necessary information about variant types. The inclusion of configuration files is crucial as they hold information about the model origin and variant types, underscoring the importance of providing these files when sharing models to ensure accurate and reliable reproduction of results. For the remaining 2\% that did not include configuration files, it is important to note that all models should ideally have these files, if not for anything else, then to identify the parent model and variant type. Configuration errors are well-documented as a significant source of problems in software systems \citep{xu2015systems, yin2011empirical, santolucito2016probabilistic}, and their absence in these \llm releases raises concerns about potential issues during deployment. 

\textbf{Since 2022, the 52,227 \llm variant releases on \HF have all been derived from only 299 \bases. The top 15 \bases account for 85.80\% of these releases, with Llama, Bert, and Mistral leading the community in the number of times they have been adapted. However, the prevalence of these top 15 \llms relative to their age shows that Gemma and Mistral are growing faster than Llama, which has seen the most explosive development when their age is not considered. This highlights the concentrated development efforts within the community.}

\begin{figure*}[t]
\centering
\includegraphics[width=0.7\textwidth]{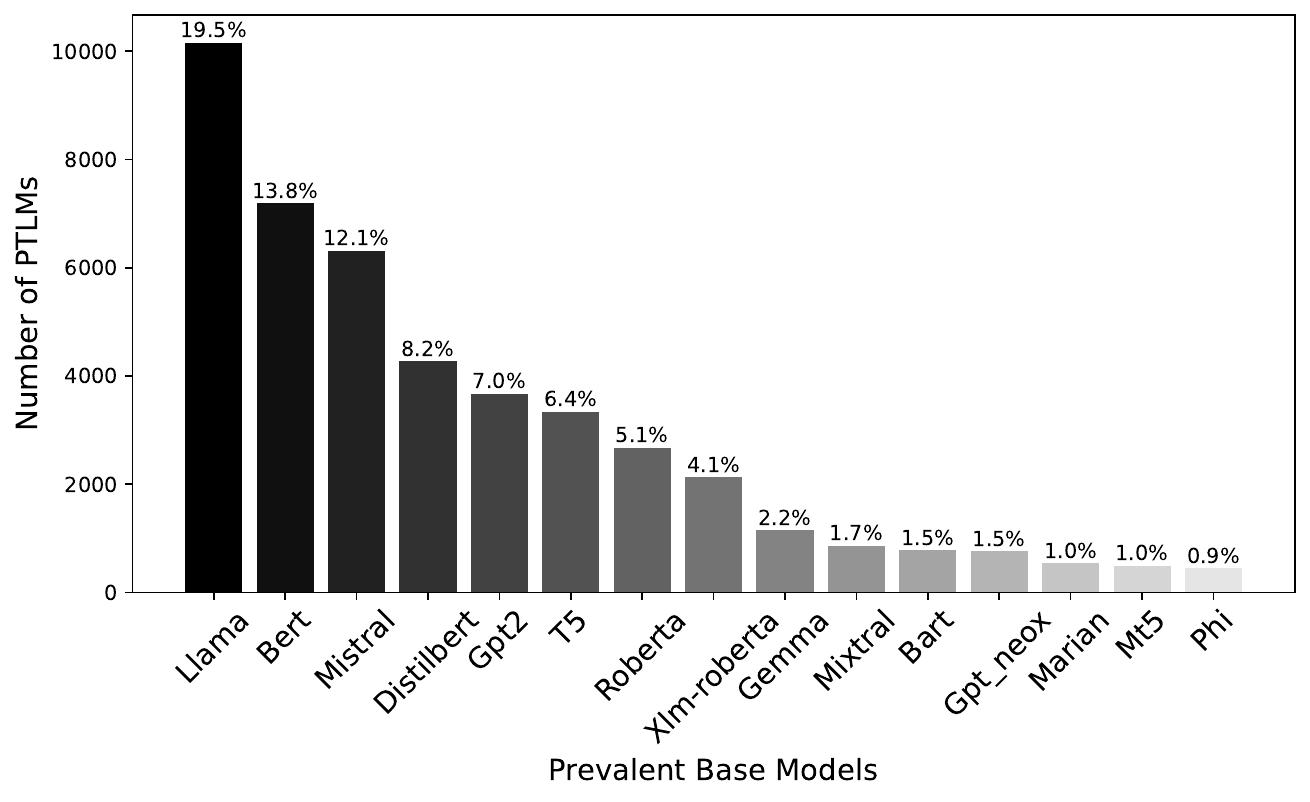}
\caption{Percentage of \llm releases by each top 15 (out of 229) base model on \HF, calculated as the number of \llms released using each base model divided by the total number of \llms studied (52,227) as of March 17, 2024, multiplied by 100.}
\label{ptlms_base}
\end{figure*}

Our analysis reveals that the initial \llm variant release using the GPT2 base occurred in March 2022. We visualize the distribution of these top 15 base models in \Cref{ptlms_base}, which contributed to 85.80\% of the studied models. It is evident that Llama, Bert, and Mistral dominate the landscape of \llm releases on \HF, contributing 19.45\%, 13.78\%, and 12.08\% of the total studied models among the top 15 \bases, respectively. Conversely, Phi, MT5, and Marian contributed the least, with 0.85\%, 0.95\%, and 1.04\% among the top 15 \bases. This concentration highlights the community's preference for Llama, Bert, and Mistral base models, likely due to their suitability for various NLP tasks.

\begin{figure*}[!ht]
\centering
\includegraphics[width=0.7\textwidth]{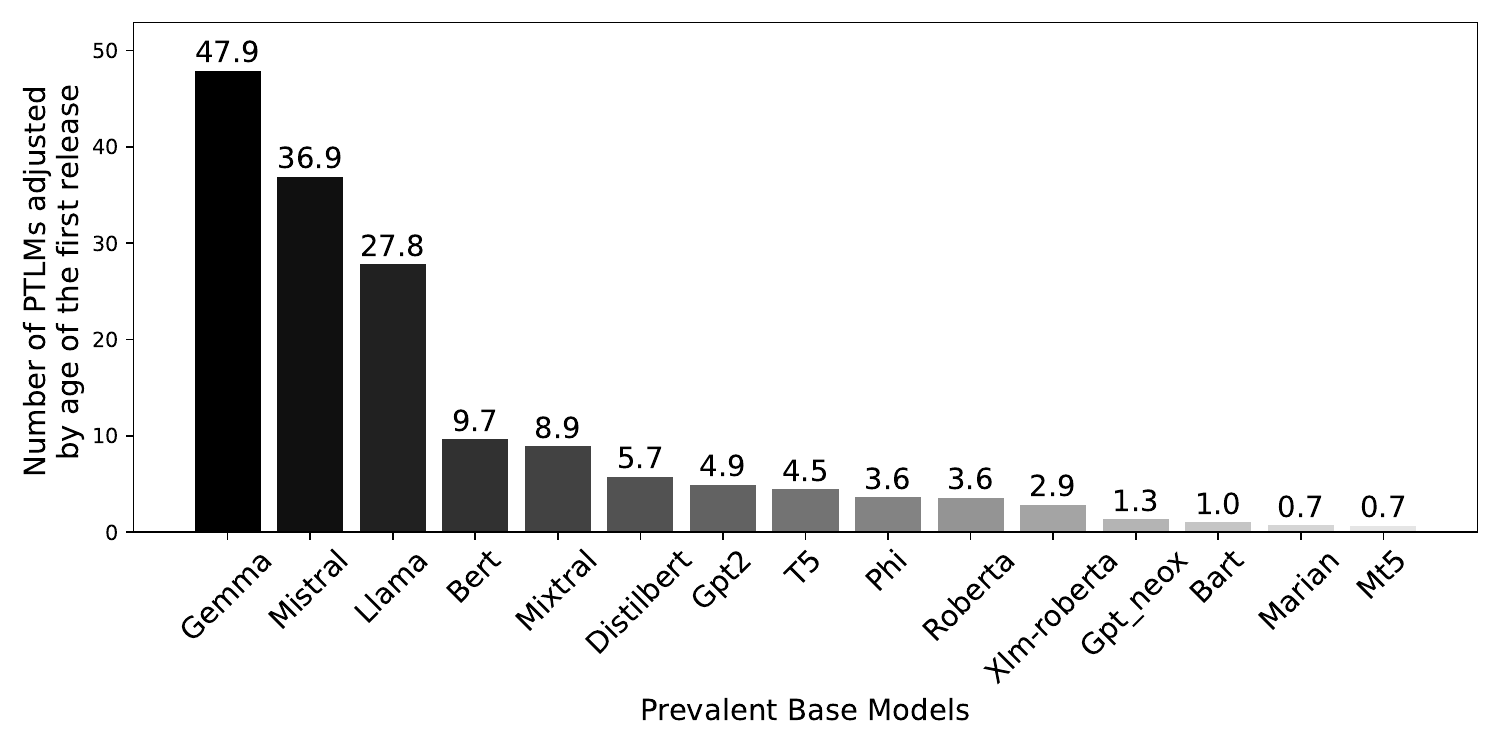}
\caption{Top 15 \bases (out of 299) as of March 17, 2024, by number of \llm variant releases on \HF per day. The number of \llms released is normalized by the age (in days) of the first \llm released using that \base on \HF.}
\label{base_model_prev}
\end{figure*}

We further explored the prevalence of each \base, adjusted by the age of its first \llm release (in terms of number of models per day), illustrated in \Cref{base_model_prev}. Gemma has seen the most explosive development, and even Mistral was growing faster than Llama, which makes them stand out prominently with the highest proportion of \llm releases per day (47.9\%) and (36.9\%), highlighting their popularity among practitioners. Conversely, \llms such as MT5, Marian, and Bart have fewer releases, indicating lower adoption rates within the \HF community by the practitioners. The concentration of development efforts around these top 15 \bases is further emphasized by their collective contribution of 85.80\% of the 52,227 \llms, illustrating the community's focus on a select group of \bases.

This observation may be due to other factors that make Gemma, Mistral, and Llama particularly appealing to \HF developers and users, such as their high performance, ease of adaptation, and broad applicability. Further research is needed to understand why these models are more popular in the community.

\noindent\textbf{Among model names, configuration files, and tags, model names and tags show diverse variant type indications. Specifically, 12.76\% of models specify these diverse variant types in their names, compared to 5.63\% in their tags.} Our analysis reveals that practitioners on Hugging Face (HF) denote the variant type of their models in three locations: model names, configuration files, and tags. Diverse variant type indications, such as `ft', `AWQ', and `deduped', are commonly specified in model names and tags. However, only the indications of models that have undergone parameter conversions, like `float16', are found in configuration files. Of the 52,227 models analyzed, 12.75\% indicated their variant types in model names, and 5.63\% did so in tags. In contrast, 19.22\% of models indicated their variant types in configuration files. Furthermore, when practitioners include this information in model names, they also tend to include it in the tags or configuration files. We observed that 2.48\% of models contained indications of variant types in all three aspects. However, 8.51\% specified this information solely in model names, 13.90\% in configuration files only, and 1.17\% in tags only. This implies that while model names are a prominent location for diverse variant type indications, tags are used less frequently and often redundantly, it highlights the variability in how practitioners choose to document variant types, with some relying exclusively on one location while others use multiple locations.

\textbf{70.72\% of 52,227 \llm releases does not state their variant type at all.} This absence of \vtype information may significantly complicate the process of selecting the most suitable model for a particular task, because different \llm \vtypes have distinct characteristics and performance metrics. For instance, a fine-tuned model might excel in a specific domain but struggle with generalizability \citep{ding2023parameter, howard2018universal}, while a distilled model might offer faster inference times but sacrifice some accuracy. Without knowing the \vtype, users may face challenges in selecting the optimal model. This can lead to inefficient exploration through trial-and-error approaches, potentially wasting valuable time and computational resources. Additionally, the lack of clear variant information may result in users independently recreating and uploading certain variants, leading to redundant work and further inefficiencies. Even if issues are identified with a base model, such as copyright concerns or performance limitations, knowing a model’s derivation and adaptation method remains crucial for compliance and legal purposes. Clear variant type labeling enhances trust and communication between developers and users. When developers provide comprehensive information about their models, including variant type, users feel more confident in their choices and are more likely to engage with the model.

\textbf{14 distinct variant type indicators are extracted from model names, configuration files, and tags.} Our analysis shows that 14 different indicators are being used by the practitioners to indicate the variant types of their model. In this case, We classified these indicators into four distinct categories: Fine-tuning, Deduplication, Quantization, and Knowledge Distillation. It is important to note that 0.7\% of models used multiple adaptation methods on a single base model, resulting in multiple variant types. In this case, we decided not to categorize them separately but maintained their variant type and duplicated the models with such multiple variant types for the analysis. This approach was taken to avoid overcomplicating the classification and to maintain clarity in our analysis.

In our examination, we found two indication of Fine-tuning: ``finetuned" and ``ft". They signify instances where models have undergone additional training to enhance their performance on specific tasks.

Furthermore, our analysis identified eleven indication of Quantization: indicating various techniques and methods used to decrease the bit-width or precision of numerical values within models:

\begin{itemize}
\item \textbf{4bit}: quantization to 4-bit precision.
\item \textbf{8bit \& q8}: quantization to 8-bit precision.
\item \textbf{Int4}: integer quantization to 4-bit precision.
\item \textbf{QAT}: Quantization-Aware Training, a technique where quantization constraints are applied during training.
\item \textbf{awq}: adaptive weight quantization, a technique where quantization is applied to model weights.
\item \textbf{float16}: quantization to 16-bit floating point precision.
\item \textbf{int8}: quantization to 8-bit integer precision.
\item \textbf{ptq}: Post-Training Quantization, a technique where quantization is applied after model training.
\end{itemize}

We also identified one indication of Deduplication: ``deduped''. It indicates instances where efforts have been made to eliminate duplicate or redundant information within models.

Our analysis found one indication of Knowledge Distillation: ``distilled", which denotes instances where models have been trained using knowledge distillation techniques to transfer knowledge from a larger model to a smaller one.

These categories provide insights into the different techniques and processes applied during the development and refinement of \llms available on HF. They also highlight the ways in which variant types are documented and where these variant types can be located. However, the fact that there are four identified variant types, and there are redundant specifications of these variant types in different aspects of the \llm repositories, shows that inconsistency in choosing a specific location for indicating these variant types implies a potential risk that many variant types might not be adequately documented. This inconsistency can lead to confusion and difficulty in identifying the specific adaptation method for a specific \llm.

The widespread use of these indications within the \HF community suggests they could serve as foundational segments for designing a mechanism for future semantic versioning of \llms. Establishing such standards could enhance clarity and interoperability in model development practices across platforms, ensuring developers and users alike have a clearer understanding of model functionalities and changes over time.

\textbf{Based on the available information, Quantized \llm releases, constituting roughly 69.3\% and Fine-tuned \llm releases, constituting roughly 29.6\% of 15,287 \llm releases, are the most released model variant types on \HF.} 

\begin{figure*}[t]
  \centering
  \includegraphics[width=0.7\textwidth]{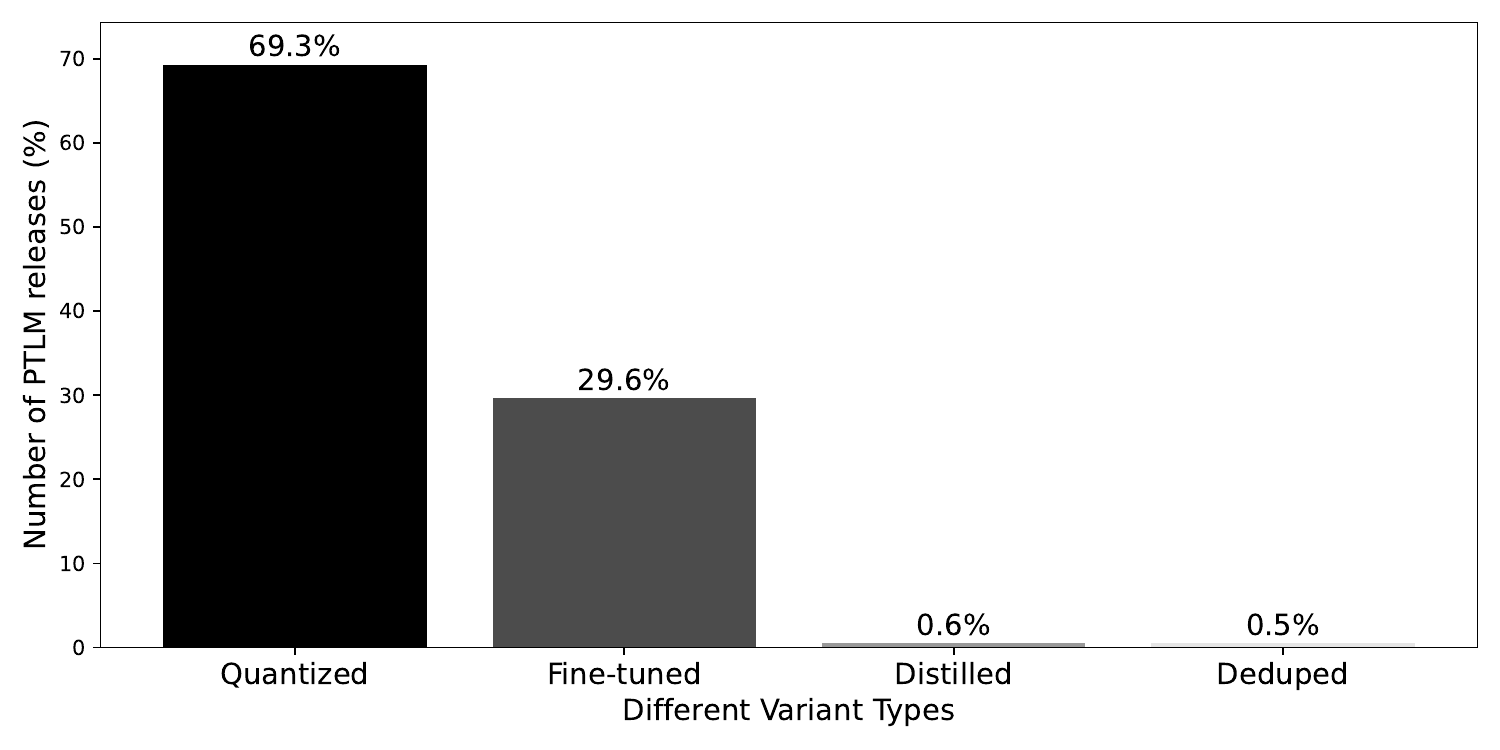} 
  \caption{Distribution of \vtypes by the number of releases out of 15,287 releases.}
  \label{base_var}
\end{figure*}

It is evident in \Cref{base_var} that the community is primarily adopting two methods of modifying the \bases: \QT and \FT. This trend could be attributed to several factors, including reduced model size through quantization, which enhances accessibility and specific use cases and easy performance improvement through fine-tuning. \QT has been optimized to reduce model size without compromising much on performance, leading to greater efficiency, while \FT has been optimized to significantly improve model performance metrics such as accuracy \citep{dettmers2024qlora, Martin2024, wortsman2022robust, liu2022few}. However, as high percentage (70.72\%) of practitioners fail to indicate the variant type of their models, it is difficult to draw definitive conclusions about the most released \llm variant types in this study.

Furthermore, based on our results, among the top 15 released \bases, Llama and Mistral are prominently used for quantization, while Bert and Distilbert emerge as the predominant choices for fine-tuning \bases. Specifically, Llama and Mistral account for 85.65\% of quantized \llms, highlighting their dominance in this technique. Similarly, 52.93\% of the fine-tuned \llms are based solely on Bert and Distilbert models, highlighting their significant role in this method. Additionally, 91.66\% of the deduplicated \llms in our study originate from the GPT-NeoX \base, illustrating its widespread adoption for this approach. Conversely, Bert and Distilbert contribute to 76.05\% of the distilled \llms, emphasizing their prevalence in this method. \\

\takeaway{Summary}{Since 2022, 52,227 \llm releases on \HF have been derived from only 299 base models, with the top 15 \bases accounting for 85.80\% of these releases. Llama, Bert, and Mistral are the most frequently adapted models, although Gemma and Mistral are growing faster relative to their age compared to Llama. Variant types are indicated in model names, configuration files, and tags, with 12.76\% of \llms specifying these variant types in their names, making it the most common method for variant type indication. However, 70.72\% of \llms do not specify their variant type. Four distinct variant type indicators—Quantized, Deduped, Distilled, and Fine-tuned—are observed. Quantized and Fine-tuned \llms are the most prevalent, constituting approximately 69.3\% and 29.6\% of releases, respectively.}

\paragraph*{Training dataset transparency on \HF} \textbf{Our automatic method for identifying \llms with training dataset information in the metadata on \HF shows that out of 52,227 \llms, 33,964 (65\%) have dataset metadata. Among these, only 24\% (8,157) explicitly indicate their training datasets in the metadata. Further manual analysis of the model cards for the remaining 25,807 \llms reveals that just 12\% mention the names of their datasets, and only 2\% provide a link to the dataset.} 

Our results show that among the \llms with dataset metadata (33,964), only 24\% specify the training dataset in the metadata. Consequently, we manually analyzed those \llms that do not specify datasets in their metadata by reading their model cards. A manual analysis of a random subset (n=379) of those lacking training data information (25,807) revealed that only 12\% mention the training dataset name in the model card, and a mere 2\% provide links to the source. This lack of transparency may hinder users' ability to assess potential biases in the training data and their impact on the model's suitability for specific tasks. It is important to note that all the datasets identified by the HfAPI are datasets hosted on \HF, while others mentioned in the model cards could be either external datasets or \HF datasets.

Our findings on dataset transparency contrast with those of \citep{pepe2024hugging}, who reported that 14\% of models identify their datasets via specific tags, and that 61\% of the top-downloaded models that they manually sampled document their training datasets. In comparison, our broader analysis of 52,227 pre-trained language models (PTLMs) reveals a significant lack of transparency. Out of our 52,227 models, 33,964 models contain dataset metadata, but only 8,157 (24\%) of the latter specify their training datasets in the metadata accessible via the Hugging Face API. For the remaining 76\% of PTLMs (25,807 models) not specifying their training dataset in \HF metadata, manual analysis of a random sample of 379 models showed that only 12\% of them mention the dataset name, with $\frac{1}{6}$ of the latter also providing a URL, while the remaining 88\% had empty data cards. These findings are consistent with those of \citep{oreamuno2024state}, who reported that 71.52\% of datasets lack a dataset card and exhibit inconsistent documentation across sections. Both studies underscore the pervasive issue of insufficient dataset transparency, which complicates the evaluation of biases and the suitability of models, regardless of their popularity.

\textbf{Only a minority of models from deduplication (24\%), fine-tuning (22.3\%), and quantization (16.4\%) were accompanied by training datasets. Even knowledge distillation, with a slightly higher rate (32.9\%), had a substantial portion without explicitly specified datasets.} 

\begin{figure*}[t]
  \centering
  \includegraphics[width=0.7\textwidth]{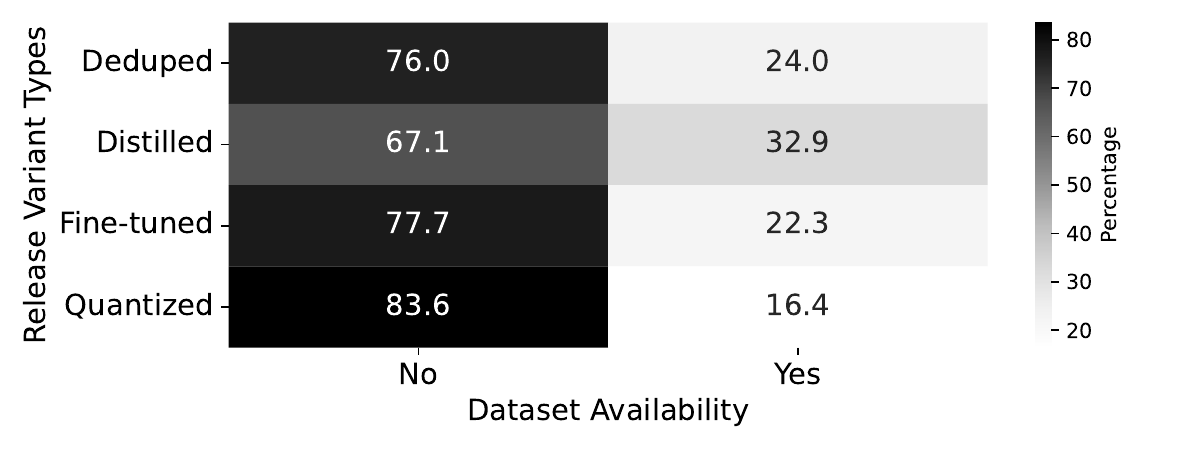} 
  \caption{Distribution of release \vtype with available training datasets from a total of 8,157 \llms.}
  \label{RQ2dataset}
\end{figure*}

The result in \Cref{RQ2dataset} shows a general lack of dataset inclusion in each variant type release. However, there is still a need to understand if there is a relationship between the variant type \llm releases and the inclusion of training datasets with them. We therefore conducted a chi-square test for each variant type to understand the relationship between the variant types and dataset inclusion. For each test, we constructed a confusion matrix to compare the presence or absence of dataset metadata with each variant type. The analysis revealed significant associations for fine-tuned releases ($\chi^2$ = 71.68, p $< 0.05$), distilled releases ($\chi^2$ = 11.62, p $< 0.05$), and quantized \llm releases ($\chi^2$ = 64.69, p $< 0.05$) with dataset availability. This suggests that there might be a tendency for uploaders of fine-tuned, distilled, and quantized models to be more likely to include the training dataset. However, the significant chi-squared result does not specify the direction of the relationship, meaning that while there is a statistically significant association, we need to examine whether the inclusion rates are higher or lower for these variant types. In this case, we found that the inclusion rates are higher for these variant types. For deduplicated \llm releases ($\chi^2$ = 1.37, p = 0.24), no statistically significant relationship was observed with dataset inclusion. This is consistent with the nature of this technique: many deduplication methods might operate as black-box procedures that do not require retraining on a dataset. Therefore, the decision to publish a training dataset for deduplicated variant releases might be more influenced by specific use cases or model complexity rather than the \vtype itself. Further research is needed to explore these potential explanations and to confirm these findings. \\

\takeaway{Summary}{A large portion (76\%, n=43,453) of studied \llms lack training dataset specification in the dedicated field provided by \HF. This transparency gap hinders user understanding, as only 12\% of these \llms with missing dataset information mention it in their model cards, and a mere 2\% provide links to the dataset source.  The statistically proven disparities in dataset transparency across different variant types (below 33\%) highlight the need for improved practices and standards in model documentation.}

\noindent \emph{Model card transparency on \HF.} \textbf{33\% of the 52,227 \llms were not released with model card documentation, hindering users' understanding and responsible utilization of \llms on \HF.} For instance, howey/electra-large-qqp\footnote{https://huggingface.co/howey/electra-large-qqp}, monologg/koelectra-base-finetuned-sentiment\footnote{https://huggingface.co/monologg/koelectra-base-finetuned-sentiment}, and msintaha/gpt2-finetuned-rocstories\footnote{https://huggingface.co/msintaha/gpt2-finetuned-rocstories} do not have any model card that shows the documentation of their training datasets, hyperparameters, or intended use cases. This lack of transparency can impede efforts to evaluate model biases and replication efforts, as users of these models rely on comprehensive model cards to make informed decisions. Furthermore, this finding shows an increase in model card documentation compared to the findings by \citep{oreamuno2024state} who found that 39.62\% of 55,280 models in \HF have a model card and \citep{taraghi2024deep} who found that 53.38\% of 239,422 models on \HF have model cards. Our analysis suggests a potential improvement in the prevalence of model cards specifically for \llms on \HF compared to \citep{oreamuno2024state} and \citep{taraghi2024deep} broader findings. This improvement might be attributed to the recent increase in community interest in pre-trained language models and the timing of our study, which captures more current trends in model documentation practices.

Our findings also resonate with \citep{bhat2023aspirations}, who identified significant gaps between the proposed best practices for model cards and their real-world usage. They proposed the DocML tool to guide data scientists in improving documentation and maintaining traceability links of pre-trained models on Hugging Face. In our study, the absence of model cards in 33\% of PTLMs highlights the need for such interventions to promote accountability and transparency in model documentation.

\begin{figure*}[t]
  \centering
  \includegraphics[width=0.7\textwidth]{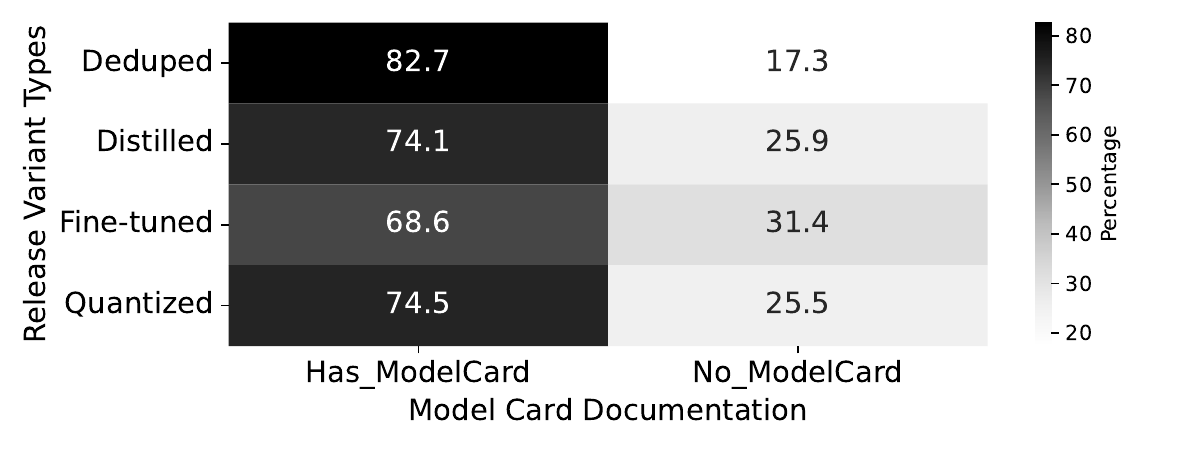} 
  \caption{Distribution of release \vtype with available training datasets from a total of 8,157 \releases.}
  \label{RQ2modelcard}
\end{figure*}

\textbf{The analysis of \llm model card transparency for the \vtypes depicted in \Cref{RQ2modelcard} reveals a significant variation in the presence of model cards across the different \vtypes. \ded models exhibit the highest percentage of release with model cards (82.7\%), followed by \qt (74.5\%) and \ds models (74.1\%). \ft models have the lowest representation (68.6\%).} Notably, over 80\% of \ded models are released with a model card, suggesting more consistent documentation practices among uploaders of this \vtype. Conversely, the lower model card presence for \ft models still shows good practices of model card releases, because 68.6\% is not too low but indicates that practitioners can do better. Improving the consistency of model card documentation for \ft models would enhance transparency, aligning their practices more closely with those of other \vtypes.

To investigate the variation, chi-square tests were conducted for each variant type to assess the relationship between release \vtypes and model card documentation. Statistically significant relationships were found for \ft models ($\chi^2$ = 43.68, p $<$ 0.05) and \qt models ($\chi^2$ = 78.52, p $<$ 0.05), indicating that the higher prevalence of model cards for these \vtypes is not by chance. However, the chi-square test for \ds models ($\chi^2$ = 0.06, p $>$ 0.05) and \ded models ($\chi^2$ = 3.53, p $>$ 0.05),  did not show a statistically significant association, suggesting that the observed percentage of model cards for the model of these variant types might be due to random chance rather than a meaningful link. Therefore, the significant relationship between model card presence and \vtype observed for \ft, and \qt models contradicts the overall lack of model card documentations reported in \citep{oreamuno2024state} and \citep{taraghi2024deep}.

Overall, this result suggests that for dataset transparency, Fine-tuned and Quantized models have the lowest dataset availability but are statistically significant in their association with the presence of datasets. For Model card transparency, Fine-tuned models have the lowest available model card documentation, while Quantized models have moderate availability. These findings underline the complexity of transparency practices across different model variants. Despite statistical significance, the practical implications of these differences highlight the need for standardized documentation practices across all \llm variants.\\

\takeaway{Summary}{67\% of \llm releases included model cards, indicating a 14\% improvement from previous findings. Despite this progress, the lack of comprehensive model cards may pose challenges for users in understanding model capabilities and limitations. This reinforces the importance of complete model cards for responsible \llm utilization. Furthermore, there's a variation in model card presence across the \llm \vtypes. \ded models have the highest documentation rate (over 82.7\%), while \ft models have the lowest (68.6\%). The observed statistical associations between model card presence and dataset transparency suggest variant-specific documentation practices. This highlights the need for improved and standardized documentation practices on HF to support clear and responsible model use.}

\subsection{\textbf{RQ-3:} To what extent do versioning identifiers in \llm names align with actual changes in \llm versions on \HF?} 

\subsubsection{Motivation}
Unlike traditional software engineering practices, where the version number of a release often reflects clear changes such as bug fixes or feature additions, the specific improvements associated with version updates in \llms remain unclear. Understanding the specific changes or enhancements made between versions, such as performance improvements or configuration adjustments, is important for informed decision-making, particularly from a user's point of view. Unclear versioning strategies lead to uncertainty about whether to risk an update or not, which is the essence of why semantic versioning practices were developed for software engineering. Therefore, this RQ focuses on the changes introduced in successive \llm releases, specifically exploring the nuances of version numbering in model releases.

\subsubsection{Approach}
\begin{figure*}[t]
  \centering
  \includegraphics[width=0.7\textwidth]{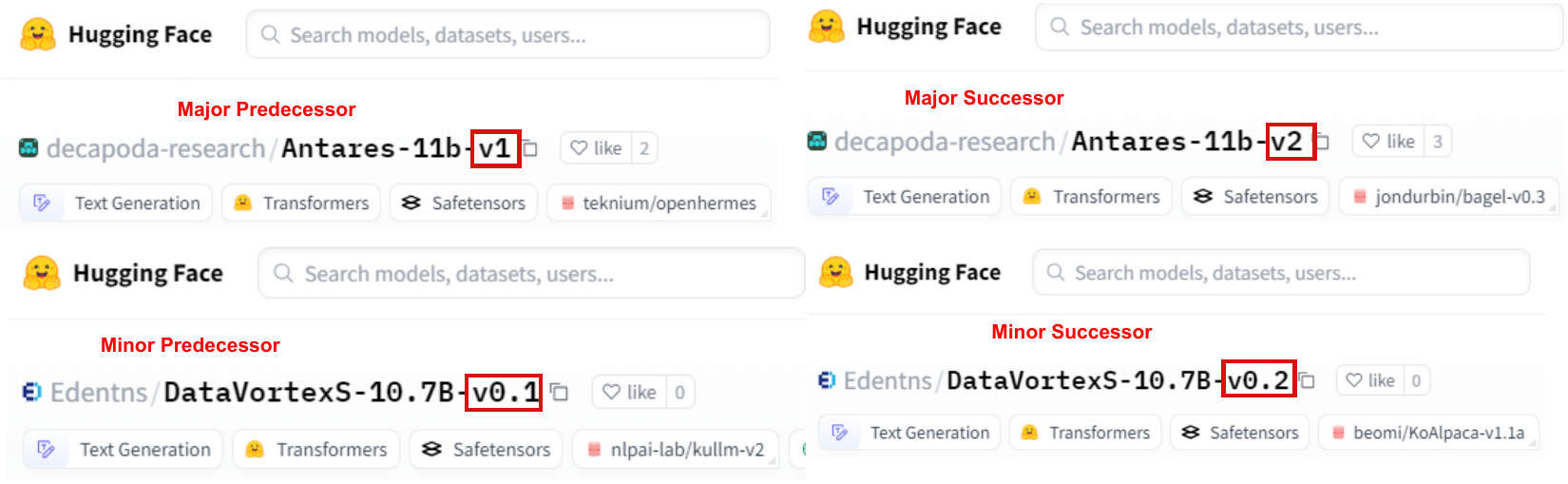} 
  \caption{Examples of predecessors and successors of \llms on \HF, randomly selected for analysis.}
  \label{successor}
\end{figure*}

\noindent\emph{Mapping of predecessors and successors of major and minor versions on \HF.} This RQ examines the differences between the current model we are studying and its predecessor (the version before the one under study) or successor (the version after the one under study) within the major and minor model version categories identified on \HF. We give examples of successors and predecessors of \llms on \HF in \Cref{successor}. This distinction is based on the two types of versioning identifiers observed in RQ1: major versioning, such as v1 and v2 and minor versioning such as v1.x and 2.x. This is to understand how major and minor versioning practices correspond to actual changes between two versions of a single \llm. This understanding will clarify the nature of modifications between versions, aiding in the assessment of versioning practices. For consistency, we focus on model names that include a segment with version identifiers, as reported in RQ1. Although models might indicate versions in different ways, such as using only numbers without the preceding 'v', we concentrated on those following the `v\textbackslash d+(\textbackslash.\textbackslash d+)*' pattern. Out of all the studied models, only 3,471 have these versions, which now serve as the basis for this analysis.

To map predecessors and successors, we employed a two-step approach. First, we selected a random sample of 330 major versions (out of 2,329) and 288 minor versions (out of 1,142) with a 95\% confidence level and a 5\% margin of error. Next, we manually identified the predecessors and successors of these samples by accessing the repositories of each selected version to locate their corresponding predecessor or successor based on their names under the same ownership. We ensured that the immediate predecessor or successor of the treated version was selected. For example, we identify the model version \textit{TheDrummer/Llama-3SOME-8B-v1}\footnote{https://huggingface.co/TheDrummer/Llama-3SOME-8B-v1}, which is a predecessor to another model. By inspecting the owner's repository, we identified its immediate successor, \textit{TheDrummer/Llama-3SOME-8B-v2}\footnote{https://huggingface.co/TheDrummer/Llama-3SOME-8B-v2}. In summary, for major versions, we ensured that if a model in our study was v1 (predecessor), we selected v2 (successor) from the owner's repository to form a pair. Conversely, if a model was v2 (successor), we selected v1 (predecessor) to review previous releases and understand the differences between the current release and its predecessor. If v2 was unavailable for a v1 model, we selected v3. Similarly, if v1 was unavailable for a v2 model, we selected v3. For versions that are not the first, such as v2, we prioritized their predecessor, and if the predecessor was not found, we took the successor. In essence, we selected only one version, either the predecessor or successor, of the version we were studying for pairing. However, for versions that are the first, we prioritized their successor. The goal was to understand the differences between two paired versions of \llms. In cases where neither the immediate predecessor nor the immediate successor versions were available, we selected any other available version of that model, whether predecessor or successor. We applied the same approach to minor versions.

\noindent\emph{Determining the differences between the model artifacts of predecessors and successors in major and minor versions.} After completing the manual mapping procedure, we calculated the percentage of successfully paired versions of \llms. These identified pairs of predecessors and successors were then used for the remaining analysis in RQ3. Although replacing models that do not have both a predecessor or successor is an option, this approach is time-consuming due to the manual nature of identifying these relationships. Furthermore, by focusing on models with confirmed predecessors or successors, we aim to accurately assess versioning practices and ensure data integrity. This approach offers a clear overview of practitioners' consistency in versioning and helps us evaluate whether they are maintaining appropriate versioning practices on \HF.

After mapping the predecessors and successors in both major and minor versions, we explored the differences between the model artifacts of these versions on \HF. Model artifacts here refer to the base model (the initial model architecture and parameters), model binary file size (indicating storage requirements), model binary file pointer size (affecting memory usage), model readme size and content (overview and usage instructions), and model card content (detailed documentation about the model's use, limitations, and ethical considerations). Understanding these changes is important because they help us determine if there are actual modifications between the predecessor and successor versions, especially to the model weight, which is the main attribute indicating a substantial model change on \HF.

We assume that if there are no changes to the model weight file size between the predecessor and successor of a major or minor version, the model owner may have uploaded the model to a new repository without making any substantive changes. This assumption helps us avoid unnecessary exploration of such models. However, if there are changes to the model weight binary file, it is important to explore the documentation to understand what actually changed.

To achieve this goal, we developed a custom Python script \citep{SAILResearch2024} that extracted the base model, model binary file size, model binary file pointer size, model readme size, and model card content. To find the differences between the base models of predecessors and successors, we made a direct comparison of the base model names. For the model binary files, readmes, and model weight file pointers, we compared their sizes. This approach helped us understand if there were modifications between the versions.

For model cards, we assessed content similarity using a word tokenizer from the NLTK\footnote{https://pypi.org/project/nltk/} library and a SequenceMatcher from the difflib\footnote{https://docs.python.org/3/library/difflib.html} library. NLTK is a well-established NLP library that offers functionalities like tokenization, stemming, tagging, parsing, and semantic reasoning. Difflib, on the other hand, provides tools for comparing sequences of words or characters, finding similarities, and calculating differences between them. In this context, we used Difflib to compare the sequences of tokenized words from the model cards of predecessor and successor versions, allowing us to evaluate the extent of content changes between them. Our script was designed to determine if there are changes to the model card content or not.

After determining the differences in the model artifacts of predecessors and successors in major and minor versions, it is important to understand if these differences are statistically significant. To calculate statistical significance, we utilized Fisher's exact test \citep{boslaugh2012statistics}. Fisher's exact test is a statistical measure used as an alternative when chi-square tests are invalid due to low expected frequencies \citep{williams2019chapter}. The imbalanced distribution of the outcomes in predecessor and successor mapping for major and minor versions justified the use of Fisher’s exact test instead of the chi-square test.

\noindent\emph{Determining the changes in the predecessors that lead to the deployment of successors.}
Upon discovery of differences in the model weight files and model cards between predecessors and successors in major and minor versions, we employed a manual analysis approach to understand what changed between these versions. This involved a thorough examination of the model cards and Readmes documents associated with each version where the model weight file size had changed, along with the Readmes and model card contents. While configuration files hold valuable information regarding changes, we prioritized model cards and Readmes documents because they typically provide a more comprehensive overview of the model's details and updates.

Subsequently, the two authors independently reviewed and labeled observations from all model cards and Readmes of predecessors and successors where the model weight had changed. For example, after comparing the Readmes content of the predecessor \textit{nitky/Superswallow-70b-v0.2}\footnote{https://huggingface.co/nitky/Superswallow-70b-v0.2} with its successor \textit{nitky/Superswallow-70b-v0.3}\footnote{https://huggingface.co/nitky/Superswallow-70b-v0.3}, we found no changes other than the model name. In such cases, we labeled the scenario as ``No change." However, when slight changes were observed in the model cards, similar to modifications in YAML configuration, we labeled these changes accordingly. Notably, such changes increased the model weight file size of the later version. It is important to note that Readmes content on \HF includes both metadata and model card information. Therefore, we compared metadata for the Readmes content and model card details separately.

Following the review of Readmes and model cards, all observed changes were systematically categorized. Initially, the first author identified ten distinct categories based on the information available on \HF, derived from an initial manual review of 50 models conducted in RQ1. For example, details such as batch size, token size, and hidden layer size were categorized under ``Configuration" because they are typically found in the configuration files. Similarly, any information related to datasets was categorized under ``Dataset." After forming these preliminary categories, the first author proceeded to classify the observed changes within these categories. This initial classification was then independently reviewed by the second and third authors to ensure accuracy and consistency. They cross-checked the categories against the observed changes to verify their relevance and appropriateness.

During this review process, two categories—``Language" and ``File Format"—raised concerns. Specifically, the first author had categorized the abbreviation `en' under ``Language" and `GGUF' under ``File Format." To avoid creating too many classifications and for the sake of simplicity, it was decided after thorough discussions among the first three authors to scrap the ``Language" and ``File Format" categories. Instead, these items were reclassified under a newly created ``Other" category. This collaborative review and reclassification process ensured that all observed changes were accurately categorized, reflecting the varied nature of the information present in the model artifacts while maintaining a manageable number of categories.

We further conducted a statistical significance test of the distribution of these categories across major and minor versions using Fisher's exact test, with Bonferroni correction applied for multiple comparisons. To understand if there is an association between these categories, i.e., whether a change in one category leads practitioners to make changes in another category, we used the phi coefficient. Phi is a measure of association between two binary (nominal) variables, specifically used in 2x2 contingency tables to quantify the degree of association between the variables. Additionally, the phi coefficient can be interpreted as an effect size, indicating the strength of the relationship between the two variables. We then interpreted the phi coefficient following \citep{akoglu2018user}, where $\phi > 0.25$ indicates a very strong relationship, $\phi > 0.15$ a strong relationship, $\phi > 0.10$ a moderate relationship, $\phi > 0.05$ a weak relationship, and $\phi > 0$ indicates no or a very weak relationship, to examine the relationships between changes in different categories.

\subsubsection{Results} 
\noindent\emph{Mapping of Predecessors and Successors of major and minor versions.}
\textbf{57\% of major and 65\% of minor version releases are missing from the model owner's repositories.} Despite the large number of models hosted on the \HF repository, continuity between versions is disrupted due to practices such as utilizing separate repositories for each release instead of maintaining a single repository for all versions. This fragmentation can lead to situations where predecessor versions are missing, as seen with 9.92\% of the 141 successfully mapped major versions. For example, models like \textit{Sandrro/text\_to\_function\_v2}\footnote{https://huggingface.co/Sandrro/text\_to\_function\_v2} lack a v1 counterpart despite having 10 subsequent releases of the same \llm variant and are not found anywhere on \HF at large. Similarly, major versions like \textit{yacine-djm/binary\_v4}\footnote{https://huggingface.co/yacine-djm/binary\_v4} sometimes lack direct predecessors or successors despite having 24 releases under the same owner, indicating possible removals from the repository. This practice can inconvenience users needing access to prior or subsequent versions of their models. We found 57\% of 330 major \llm versions in this scenario. Similar observations apply to minor versions, exemplified by \textit{lmsys/vicuna-33b-v1.3}\footnote{https://huggingface.co/lmsys/vicuna-33b-v1.3}, which has no predecessor or successor despite up to 18 \llms in the owner’s profile. We also found 65\% of minor \llm versions in this scenario.

\noindent\emph{Difference between the model artifacts of predecessor and successor of major and minor versions} \textbf{Our analysis results show that model cards (71\%) experience the most changes between major versions of \llm variants on \HF, comparing directly between versions such as v1 and v2, regardless of intermediate releases like v1.1 or v1.2. In contrast, base models (13\%) experience the least changes.}
We found that the \base changed in 13\% of the analyzed cases, indicating that only a few versions with major identifiers alter the underlying architecture. However, model weight files changed in 67\% of the cases, suggesting substantial updates to the version's learned parameters. Additionally, model weight file pointers were modified in 50\% of the models, reflecting adjustments in how the model version weights are referenced. Furthermore, README files were updated in 68\% of the models, highlighting frequent revisions to usage documentation. Finally, model cards were modified in 71\% of the cases, indicating significant updates to the descriptive metadata.

\textbf{Our further analysis of minor versions shows that model cards (80\%) still experienced the most changes between two different releases of the same \llm variant on \HF, while the model's base model was not changed in these versions.}
This finding suggests that updates in the predecessors and successors of minor versions typically retain the same architecture. Model weight files were altered in 48\% of the cases, indicating less frequent but still significant parameter updates. Model weight file pointers changed in 40\% of the cases, showing moderate adjustments. model cards were modified in 80\% of the cases, indicating regular updates to the documentation, while README files were updated in 75\% of the model versions, reflecting ongoing refinements to the metadata.

\textbf{The observed differences in the attributes of major and minor versions are only statistically significant in model weight files and base models.} We statistically compared the observed differences in the model artifacts of predecessors and successors between major and minor versions. It is evident that only the differences in the base model artifacts and model weight file attributes of \llm releases in major and minor versions are statistically significant. No significant differences were found for model binary file pointers, README files, or model card content across major and minor versions. \Cref{statistic} presents the p-values, odds ratios, and confidence intervals indicating the statistical significance between major and minor model artifacts. Specifically, there is a statistically significant (p $<$ 0.05) difference in the base model artifact between major and minor versions, with a large odds ratio (inf) and a wide confidence interval (1.855 to 522.799). The odds ratio has such values due to no observations of \llm releases in minor versions changing their base model. Similarly, the differences in the model weight attribute of \llm releases in both major and minor versions are statistically significant (p $<$ 0.05) with an odds ratio of 2.23, indicating that changes in model weight are 2.3 times more likely in major versions than in minor versions, with a confidence interval between 1.32 and 3.79.

\renewcommand{\thetable}{4}
\begin{table}[t]
\centering
\caption{Statistical difference between the changes in major and minor releases.}
\label{statistic}
\begin{tabular}{lrrr}
\toprule
\textbf{model artifacts} & \textbf{P-values} & \textbf{Odd Ratio} & \textbf{Confidence Interval} \\
\midrule
\base & $<$0.05 & inf & 1.855 - 522.799 \\
Model weight file & $<0.05$ & 2.2373 & 1.321 - 3.790 \\
Model weight file pointer & $>$0.05 & 1.5214 & 0.906 - 2.556 \\
Readme & $>$0.05 & 0.71111 & 0.400 - 1.263 \\
Model Card & $>$0.05 & 0.6097 & 0.331 - 1.122 \\
\bottomrule
\end{tabular}
\end{table}

These findings suggest that major versions, as expected, are more likely to include substantial changes, such as alterations to the base model and model weights. This aligns with the principles of semantic versioning, where major versions typically involve significant changes that may affect compatibility, while minor versions involve incremental improvements and fixes that maintain compatibility. The lack of significant differences in other artifacts like model binary file pointers, README files, or model card content across major and minor versions indicates that these segments are less likely to be altered significantly between versions, which aligns with the idea that documentation and pointers often receive more consistent updates across all types of releases.\\

\takeaway{Summary}{57\% of major versions and 65\% of minor versions of \llms on \HF lacked versioning continuity as a result of missing versions, leading to ambiguities for users needing previous versions or updates.  Similarly, while 87\% of models maintain the same \base across versions, 13\% switch to entirely different ones. Changes are frequently observed in model weight files (67\%), pointers (50\%), readmes (68\%), and model cards (71\%) in major versions, while minor versions show changes in 48\%, 40\%, 80\%, and 75\%, respectively. However, these changes are only statistically significant in the \base and model binary files, with substantially larger odds ratios (inf and 2.237) falling within the 1.855 to 522.799 confidence interval.}

\begin{figure*}[t]
\centering
\includegraphics[width=0.8\textwidth]{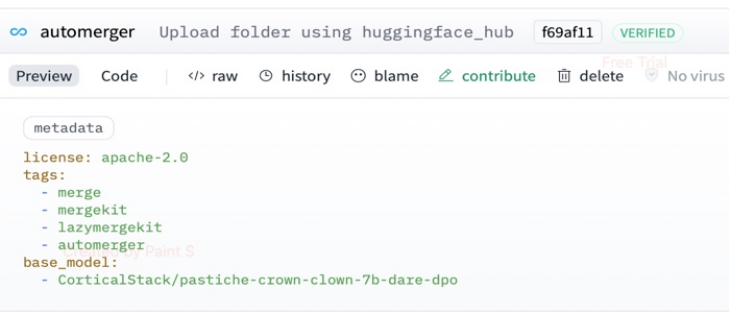}
\caption{The README contains content that differentiates it from the model card content. While the model card does not include the metadata found in the README, the README sometimes also contains all the information present in the model card. Note that it is possible for a model to contain little information in the README and no information in the model card.}
\label{readme_diff}
\end{figure*}

\paragraph*{The actual changes between the predecessors and successors in major and minor versions that translated to the new version of \llms.}
Although there are changes in the model artifacts of predecessors and successors in major and minor versions of \llms, these changes are not statistically significant except for the model weight file and base model. Changes in the model weight file translate to new versions of \llms. Therefore, to understand the changes made to these major and minor versions that caused a change in the model weight files, we focused on major and minor versions where the model weight file changed, along with the corresponding model card and README. These activities resulted in the analysis of 40 README files and 40 model cards for minor versions, as well as 71 README files and 71 model cards for major versions. \Cref{readme_diff} highlights the differences between README files and model cards, which is why we consider them as separate documentations.

\textbf{Our investigation identified change patterns for major releases compared to minor releases. Major version updates typically exhibit a broader range of modifications, encompassing an a total of 28 unique changes, while minor version updates show an average of approximately 8 changes.} We categorized these changes into nine different categories. \Cref{description} presents these categories along with the frequency of changes, displayed as percentages, and the types of version in which the changes occurred. We also depict the meaning of these change types in \Cref{definition}.

It is evident that configuration, model architecture, energy consumption, performance changes are specific to major versions. Conversely, all changes identified in minor versions were also observed in major versions, indicating consistency across updates despite differing version identifiers.

\renewcommand{\thetable}{5}
\begin{table}[t]
\centering
\caption{The outcome of different activities between the predecessor and successor of major and minor version of releases on HF.}
\label{description}
\begin{tabular}{llrcc}
\toprule
\textbf{Categories} & \textbf{Changes} & \textbf{\% of change} & \textbf{Major} & \textbf{Minor}\\
\midrule
& Batch Size & 3.82 & \checkmark &  \\
& Tokenizer Version & 3.82 & \checkmark & \\
& Evaluation Metrics & 0.64 & \checkmark & \\
& Normalized Layer & 0.64 & \checkmark & \\
Configuration change (20.35\%) & Embedding Size & 0.64 & \checkmark & \\
& Hyperparameters & 3.18 & \checkmark & \\
& Number of Epoch & 4.46 & \checkmark & \\
& Hidden Layer Size & 1.27 & \checkmark & \\
& Token Size & 1.91 & \checkmark & \\
\hline
& Merged \base & 4.46 & \checkmark &\\
Model architecture change (14.01\%) & Base Model & 5.73 & \checkmark & \\
& Training Variant & 3.82 & \checkmark & \\
\hline
License change (7.64\%) & License & 7.64 & \checkmark & \checkmark\\
\hline
& Performance & 11.46 & \checkmark & \\
Performance change (14.65\%)  & Result & 2.55 & \checkmark & \\
& Evaluation \base & 0.64 & \checkmark & \\
\hline
Dataset change (14.65\%) & Dataset Specification & 10.19 & \checkmark & \checkmark\\
& Dataset Versions & 4.46 & \checkmark & \\
\hline
& Tokenizer Version & 3.82 & \checkmark & \\
Training Library change (13.37\%) & Transformer Version & 7.64 & \checkmark & \checkmark\\
& Libraries & 1.91 & \checkmark & \\
\hline
Energy consumption (1.27\%) & CO2 Emission & 1.27 & \checkmark & \\
\hline
Performance metrics change (3.82\%) & Metrics & 3.82 & \checkmark & \checkmark\\
\hline
& Task & 3.82 & \checkmark & \checkmark\\
& Tags & 3.82 & \checkmark & \checkmark\\
Other change (14.65\%) & File Format & 0.64 & \checkmark & \\
& Language & 6.37 & \checkmark & \checkmark\\
& Model Name & 4.46 & \checkmark & \checkmark\\
\bottomrule
\end{tabular}
\end{table}

\renewcommand{\thetable}{6}
\begin{table}[t]
\centering
\caption{Definition of the different activities between the predecessor and successor of major and minor version of releases on HF.}
\label{definition}
\begin{tabular}{lp{12cm}}
\toprule
\textbf{Changes} & \textbf{Description} \\
\midrule
Batch Size & Data processed per iteration during training or evaluation. Changed in newer model.\\
Tokenizer Version & Version of tokenization mechanism for NLP models. Updated in newer model.\\
Evaluation Metrics & Measures assessing model performance. Included or expanded in newer model.\\
Normalized Layer & Neural network layer standardizing input data. Adjusted or increased in newer model.\\
Embedding Size & Dimensionality of vector space for words or tokens. Increased or decreased in newer model.\\
Hyperparameters & Configurable parameters influencing model behavior. Specified or optimized in newer model.\\
Number of Epoch & Number of passes through dataset during training. Increased or decreased in newer model.\\
Hidden Layer Size & Number of hidden layers in neural network. Increased or reduced in newer model.\\
Token Size & Length or size of tokenized input sequences. Increased or decreased in newer model.\\
\hline
Merged \base & Creation of a more powerful model by combining multiple pre-trained \bases into a single entity. The number increased or decreased in the newer version. \\
Base Model & Primary model used as a foundation. Changed in newer version.\\
Training Variant & Variant of \base used for training. Changed in newer version.\\
\hline
License & Legal terms governing the use and distribution of the model. Added, changed or removed from the newer version. \\
\hline
Performance & Measure of the model's effectiveness or efficiency in accomplishing tasks. Increased or decreased in the newer version. \\
Result & Contain the evaluation result of the model's predictions or computations. Added or removed from the newer version. \\
Evaluation \base & The baseline model compared with the newer version. Changed in the newer version. \\
\hline
Dataset Specification & Detailed description or requirements for a dataset. Added or deleted in the newer version. \\
Dataset Versions & A specific iteration or snapshot of a dataset used for training or evaluating the released model. Earlier or later version used in the newer version. \\
\hline
Tokenizer Version & Version of the tokenization tool used to process text data. \\
Transformer Version & Version of the transformer model architecture employed in the model. Earlier or later version used in the newer version. \\
Libraries & The specific library or modules used for training the released model. Added or deleted from the newer version. \\
\hline
CO2 Emission & The total amount of carbon dioxide (CO2) emissions generated throughout the training. Increased or decreased in the newer version. \\
\hline
Metrics & Performance metrics used to measure the effectiveness and performance of models. Added or deleted in the newer version.\\
\hline
Task & Specific objective or goal the model is designed to accomplish. Changed in the newer version. \\
Tags & Label or identifier assigned to categorize or classify data release information. Added or removed from the newer version. \\
File Format & A specialized binary file format designed for efficient storage and inference of model. Changed in the newer version. \\
Language & The primary language(s) the model is trained on and designed to work with. Added or removed in the newer model. \\
Model Name & Unique identifier assigned to the model. Changed in the newer version. \\

\bottomrule
\end{tabular}
\end{table}

\textbf{There is no statistically significant difference between the prevalence of changes that actually occurred between the major and minor versions, except for changes that occur in the configuration, license, and others.} 

A statistical analysis of the prevalence of the nine main categories of changes between major and minor \llm releases, as depicted in \Cref{group_stat}, reveals that six out of nine categories of changes exhibited p-values greater than 0.05. These results suggest no statistically significant difference between major and minor releases for these categories. Conversely, license, configuration, and other changes showed a statistically significant difference between major and minor releases (p-value $<$ 0.05).

While license, configuration, and other changes showed statistical significance, these findings alone do not provide conclusive evidence that the current major-minor versioning practice on \HF consistently adheres to semantic versioning principles. Semantic versioning is intended to communicate the significance of changes through version numbers, with major versions denoting breaking changes, minor versions indicating backward-compatible feature additions, and patch versions representing backward-compatible bug fixes.

\renewcommand{\thetable}{7}
\begin{table}[t]
\centering
\begin{tabular}{  l |  c }
\hline
\textbf{Categories} & \textbf{p-value} \\ 
\hline
Configuration & 0.02 \\ 
Model Architecture & 1.00 \\ 
License & 0.00 \\ 
Performance & 0.42 \\ 
Dataset  & 1.00 \\ 
Training Library  & 1.00 \\ 
Energy consumption  & 1.00 \\ 
Performance metrics  & 1.00 \\ 
Other Changes  & 0.01 \\ 
\hline
\end{tabular}
\caption{Statistical comparison of changes in major and minor version of \releases.}
\label{group_stat}
\end{table}

The statistically significant differences observed in configuration, license, and other changes suggest that these aspects are prioritized or more rigorously addressed within the update practices on HF. Configuration changes, such as adjustments to batch size or tokenizer versions, are important as they directly impact model performance and compatibility with existing applications. Similarly, license updates are significant as they may introduce new usage terms or legal implications for users and developers.

In contrast, categories like model architecture, performance metrics, and dataset changes did not show statistically significant differences between major and minor releases. While these categories are important for understanding model capabilities and performance, their consistent lack of significant distinctions between major and minor updates suggests that their versioning on \HF may not fully align with the clear distinction of major and minor changes advocated by semantic versioning, or it may simply indicate that changes in these areas are less common.

\begin{figure*}[t]
  \centering
  \includegraphics[width=0.7\textwidth]{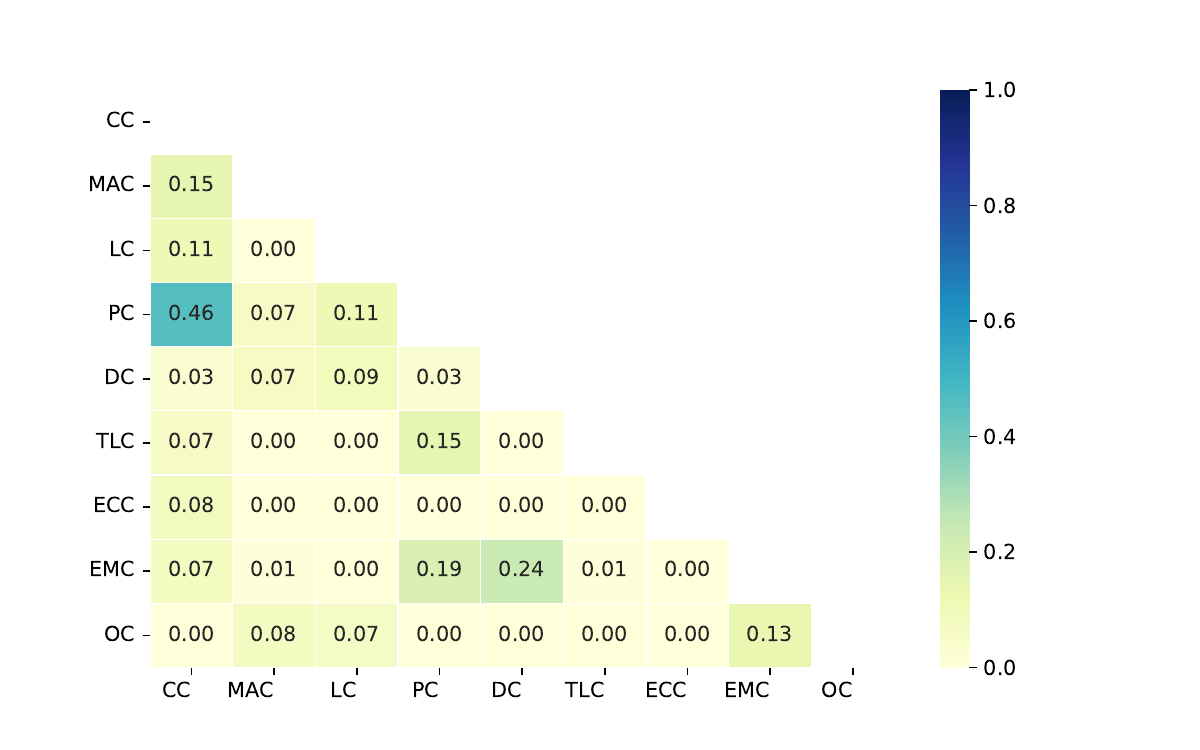} 
  \caption{Associations between several categories of changes on \HF. The acronyms are as follows: CC: Configuration Changes, MAC: Model Architecture Changes, LC: License Changes, PC: Performance Changes, DC: Dataset Changes, TLC: Training Library Changes, ECC: Energy Consumption Changes, EMC: Evaluation Metrics Changes, OC: Other Changes.}
  \label{correlation}
\end{figure*}

\textbf{The result in \Cref{correlation} shows one very strong association, four strong associations, and three moderate associations between pairs of change categories in \llms.}

We observed a very strong $\phi$ association between Performance changes and Configuration changes ($\phi$ = 0.46). When developers modify configuration settings in \llms, they are also likely to make corresponding changes in performance results. This association indicates that adjustments to configurations often co-occur optimizations or adjustments in performance metrics, ensuring compatibility and enhanced performance in subsequent versions of \llms.

There is a strong association between Evaluation metrics changes and Dataset changes ($\phi$ = 0.24), Evaluation metrics changes and Performance changes ($\phi$ = 0.19), Training library changes and Performance changes ($\phi$ = 0.15), and Model Architecture changes and Configuration changes ($\phi$ = 0.15). These strong associations suggest that updates to evaluation metrics or datasets typically coincide with adjustments in performance metrics. Similarly, changes in training libraries are closely linked with modifications in performance outcomes, reflecting the adoption of more efficient or advanced libraries to enhance model capabilities. Adjustments in model architecture often necessitate corresponding changes in configuration settings to ensure compatibility and optimal performance.

Moderate associations were observed between Other changes and Evaluation metrics changes ($\phi$ = 0.13), License changes and Configuration changes ($\phi$ = 0.11), and License changes and Performance changes ($\phi$ = 0.11). These moderate associations indicate that while modifications in evaluation metrics or licensing terms occasionally prompt adjustments in configuration or performance metrics, the relationship is less consistent compared to the stronger associations observed.

All other relationships showed weak associations ($\phi <$ 0.10), indicating that changes in these categories occur relatively independently of each other.

The observed associations between changes in different categories during \llm development provide valuable insights into how modifications are interrelated. Strong and very strong associations suggest that practitioners often alter multiple related aspects simultaneously. This pattern highlights the importance of implementing structured versioning practices on \HF to accurately capture these interdependencies. In the context of semantic versioning, such associations can significantly impact versioning practices. For instance, if a major change affects multiple related aspects of the model, it is crucial to reflect this in the version number. By rigorously applying semantic versioning, practitioners can better communicate the scope and impact of changes, improving transparency and predictability for users. Major version updates should signal substantial modifications that may require user adjustments, while minor updates can indicate incremental improvements that maintain compatibility.

Similar patterns of change associations and their impact on versioning have been well-documented in regular source code projects \footnote{Semantic Versioning 2.0.0: \url{https://semver.org}}. In software projects, semantic versioning (SemVer) practices are crucial for managing and communicating changes effectively. By drawing parallels between these practices and \llm versioning, it becomes evident that structured versioning based on observed change associations can enhance clarity and user understanding in \llm deployments. This approach not only aligns with established versioning practices but also provides a clearer framework for managing complex changes in \llms.\\

\takeaway{Summary}{We observed a total of 28 different change types in major and minor versions of \llms, categorized into nine different groups. Major version updates exhibit all 28 change types, while minor version exhibit 8 change types. There is no statistically significant difference between the changes observed in major and minor versions, except for configuration, license, and other changes.}

\section{Discussion and Implication} \label{discussion}
\subsection{Discussion}
The ubiquity of pre-trained models is on the rise, sparking interest in model registries and dedicated platforms for hosting these models. One of the major categories of these pre-trained models is \llm, which enhance various language understanding tasks and are widely acknowledged \citep{sarzynska2021detecting}. The free and open-source availability of these \llms has led to increased adoption, which has in turn brought about different challenges associated with their release and version management. As a result, we explored 52,227 \llms to answer three research questions aimed at understanding the release characteristics of these \llms, using \HF as the case study due to its status as the largest pre-trained model registry \citep{jiang2022empirical}. Therefore, this section discusses the main findings of our empirical investigation into the naming and versioning conventions, transparency of these releases, variant types, and changes that occur between different versions of \llms on \HF, aiming to improve the release process of these models.

In RQ$_1$, our analysis reveals 148 distinct naming conventions on \HF, highlighting the diverse approaches developers employ in labeling their models, reflecting individual preferences and project-specific needs. This finding supports Jiang's study \citep{jiang2023exploring}, which surveyed 108 \HF users and revealed a disparity between desired and actual naming practices. To mitigate ambiguity and inconsistencies observed in \llm names, practitioners are encouraged to adopt a structured naming convention that includes essential segments like base model and size. Our study highlights that models incorporating these details tend to receive higher downloads, suggesting increased adoption rates and facilitating better decision-making among users, particularly concerning environmental impact considerations \citep{bender2021dangers}. 

Furthermore, the disappearance of 0.2\% of \llms from the repository suggests model owners may have either deleted or renamed them. This could disrupt ongoing projects or workflows that rely on these models. To prevent such disruptions, similar to the issues seen during the left-pad incident\footnote{https://en.wikipedia.org/wiki/Npm\_left-pad\_incident}, \HF administrators should implement a mechanism that makes it difficult for practitioners to unpublish or delete an already published model. This measure would help ensure the stability and reliability of the repository for all users. 

Additionally, our analysis reveals a significant gap between declared versions (3,471) and potential versions inferred from changes (524,419) in model binary files, highlighting discrepancies in current versioning practices versus actual development frequencies. It is crucial for practitioners to diligently version every significant release to mitigate risks akin to those encountered with the Therac-25 radiation therapy machine in the late 1980s\footnote{https://en.wikipedia.org/wiki/Therac-25}. These incidents resulted in fatal radiation overdoses due to software bugs introduced in a new version. Inadequate testing and poor version control practices allowed the faulty version to be deployed to machines in hospitals. Failure to version their \llms could lead to similarly severe consequences, as applications relying on these models may be significantly impacted by untracked changes, potentially resulting in catastrophic outcomes. 

The average of 12 changes per model binary file further highlights this inconsistency, potentially misleading users relying on version numbers for assessing model updates. The presence of different model storage format compounds this issue, highlighting the importance of clear and consistent versioning to ensure interoperability and user accessibility \citep{singla2023machine, liu2020enhancing}. While HF prioritizes security and efficient storage of tensors within its ecosystem, the popularity of two main file extensions for storing model weights, ``.safetensor" and ``.bin", persists. However, this can lead to interoperability issues since other frameworks and deployment environments may not inherently support all features of ``.safetensor" and ``.bin" files. This often necessitates conversion or additional steps to integrate models into different platforms. While prioritizing security is commendable, developers should also consider adopting formats that support broader interoperability beyond the HF ecosystem.

In RQ$_2$, we identified two key areas for improvement in \llm releases on \HF: variant type transparency and training data disclosure. Despite a diverse set of 299 base models fueling a vibrant \llm ecosystem with 52,227 models, a concerning 70.72\% of releases lack explicit information about their variant type. This obscurity can hinder user understanding of specific modifications applied to a base model. Understanding the modification type is crucial, as deploying a model without knowledge of its modification type can lead to unexpected performance degradation. For example, models modified using different techniques may require specific software or hardware environments for optimal performance. Without knowing the modification type, integration into existing systems or workflows could be problematic, leading to compatibility issues or additional development costs to resolve integration challenges. Stakeholders relying on ML models, such as end-users, policymakers, or business decision-makers, often require transparency about how models are modified. Lack of clarity about modification methods can erode trust in the model's reliability, transparency, and ethical usage. Therefore, we encourage developers to consistently specify variant type information in model names, descriptions or configuration file to enhance transparency and mitigate potential risks effectively. 

Similarly, 76\% of models lack information about their training data, which is crucial as training data can significantly influence an \llm's behavior and biases. While practitioners may have valid reasons for not releasing their training datasets—such as protecting intellectual property and ensuring responsible model use—OpenAI has cited similar reasons for not disclosing the training data used in GPT-4\footnote{https://www.linkedin.com/pulse/behind-closed-doors-decision-release-training-data-gpt-4-jatasra-kr4df}. They emphasize that controlling access to the training data helps mitigate risks and ensures ethical alignment, focusing on principles like harm avoidance, fairness, and transparency. However, while security and responsible use are paramount, complete transparency, including the release of training data, remains desirable. Balancing transparency with these concerns is essential. Therefore, \HF should enforce rigorous dataset and model card documentation practices to enhance standards on their platform. This approach ensures users have access to essential information for understanding and replicating model behaviors. When practitioners use external datasets for training, they should provide links to the dataset sources, not just names, given the commonality of dataset names across the internet.

In RQ$_3$, we explored changes associated with different \llm version types on \HF. Our analysis revealed distinct usage patterns but also highlighted a significant issue: over 50\% of major and minor versions lacked clear predecessor-successor connections, primarily due to missing or deleted versions. This lack of information poses challenges for users tracking updates or locating specific \llm versions.

In addition to the issues with missing predecessor-successor connections, we found that among all model artifacts, model cards experienced the most frequent changes between major and minor versions, with 71\% of major versions and 80\% of minor versions showing updates in this area. This indicates that changes to documentation or usage instructions are the most prevalent between releases. Moreover, major versions also demonstrated significant changes in base models (13\%), suggesting potential shifts in the underlying architecture. This highlights the importance of clear versioning, as major changes can impact user workflows and require careful consideration before adoption. We identified a total of 28 different changes grouped into 9 categories. While trends in changes were observed, they did not significantly differ between major and minor versions, except for configuration, license, and other changes. This highlights the need for clear and consistent versioning practices. Without a standardized system, users face challenges such as hesitation to upgrade due to ambiguity, potential disruptions from unclear updates, and staying on outdated versions missing improvements. Therefore, semantic versioning on \HF is crucial to improve user model selection, facilitate informed decision-making, and promote responsible \llm development.

Based on our findings about the different challenges and difficulties of \llm releases on the \HF model registry, and noting the lack of a standardized notion of `model release', we define a \textit{model release as the publication of a model registry entry characterized by a meaningful name and unambiguous version identifiers, and encompassing the essential artifacts needed to successfully operate and evolve the model, such as model weights, configurations, and documentation, along with relevant provenance links to datasets and earlier model releases.} However, a significant challenge observed on \HF is that different versions often receive separate repositories instead of being included in the release history of the previous version. This practice makes it difficult to track the complete evolution of a model over time and complicates the application of semantic versioning principles, leading to fragmentation and inconsistencies in version management practices across the platform. Addressing this issue is important for reducing confusion and enhancing the clarity of and interaction with model repositories on \HF.

It is also worth noting that while our analysis focused on a diverse set of models with sufficient metadata, the distribution of model usage on Hugging Face is likely skewed \citep{osborne2024ai}. A small subset of highly popular models (e.g., the Llama, Qwen and DeepSeek families) accounts for a disproportionate share of user engagement and influence on the ecosystem. While unclear or inadequate versioning in widely used models can create significant challenges for numerous users, affecting adoption, integration, and trust in the platform, less popular models may not have the same reach, even if their versioning practices are similarly inconsistent. This skewness in usage suggests that the practical significance of our observations could vary depending on the popularity of a model. Future research could investigate how versioning practices and their consequences differ between high-impact models and those with more limited user bases.

\subsection{Towards Semantic Versioning of PTLMs}
\subsubsection{Versioning of PTLMs vs. traditional code}
The findings of this study highlight a fundamental difference between versioning for PTLMs and traditional code artifacts. In software, semantic versioning encodes backward compatibility and adherence to downstream client contracts into a concise, 1-dimensional numbering scheme \citep{semantic_2025}. This approach relies on the clear definition of 'contracts'—the expectations and compatibility between artifacts and their clients—that allows semantic versioning to systematically indicate whether a version complies with or violates clients' expectations \citep{lam2020putting}. These contracts encapsulate essential details such as API specifications, backward compatibility, functionality additions or deprecations, and bug fixes. However, PTLMs lack a direct counterpart to such contracts, i.e., there currently is no agreement on what compatibility means in the context of PTLMs. As our results demonstrate, PTLMs' inherently multidimensional nature, encompassing characteristics such as architecture, size, training data, and domain specificity, complicates the application of semantic versioning principles and concepts. Furthermore, model owners currently try to project this multidimensionality onto a one-dimensional model name, but without an established naming convention, leading to ambiguities that challenge users in assessing compatibility and understanding the implications of changes (as we have pointed out in our findings).

Unlike traditional software, PTLMs lack a standardized versioning framework due to their inherently multidimensional nature. This necessitates an approach that explicitly stores version metadata and enables automated compatibility assessments to streamline model adoption and evolution. In the following sections, we discuss (1) how to represent the 'version' of a model, (2) how to determine the right 'version' for a new model, and (3) how model version metadata can be accessed.

\subsubsection{How to Represent the ``Version" of a Model}
Representing the version of deep learning models, particularly PTLMs, presents challenges that go beyond traditional software versioning. Semantic versioning (e.g., X.Y.Z) has been pivotal in software development, helping to prevent ``dependency hell" by clearly signaling the nature of changes between versions \citep{lam2020putting, semantic_2025}. This one-dimensional structure of version numbering excels in signaling backward compatibility—where major version increments (X) indicate breaking changes, minor version increments (Y) represent new but compatible features, and patch version increments (Z) account for backward-compatible bug fixes.

The one-dimensional approach of current semantic versioning assumes that changes can be captured in a linear, incremental fashion, but this simplicity cannot capture the complexity of PTLM evolution, due to the unique dynamics of model evolution. For instance, does changing the base model imply a major version update, or does it reflect a minor change? What about fine-tuning a model with an entirely new dataset—should this be classified as a major or minor change? Furthermore, configuration and license changes—often seen in model releases—may be viewed as relatively trivial, but should they constitute patch updates, or do they warrant a more significant version change?

These uncertainties illustrate the difficulty of defining ``backward compatibility" for PTLMs. While in traditional software, backward compatibility refers to a new version not breaking existing functionality, for PTLMs, model changes such as data shifts, architecture changes, or training adjustments may not fit into the established categories of software versioning.  This makes it challenging to determine if a model is backward compatible in the same way as software, as modifications to models do not always directly correspond to ``breaking" or ``compatible" changes within the versioning system.

Therefore, PTLM versioning requires a more effective approach, one that accounts for factors beyond functional compatibility, such as dataset shifts, model architecture, task specificity, reuse methods, and training dataset modifications. For instance, a minor adjustment in hyperparameters might not necessitate a major version update under traditional schemes, but it could have significant downstream impacts.

Our study reveals that existing naming practices on Hugging Face often attempt to embed information about model changes, albeit inconsistently. Beyond compatibility, a critical aspect of semantic versioning should be the inclusion of provenance information—details about a model's origin and the transformations it has undergone. In the context of versioning PTLMs, our paper identifies specific fields that may be important for a more accurate versioning system. These include: identifier, base model, model size, and training mechanisms. By incorporating these fields, versioning could better reflect the nuances of model evolution and provide users with more transparent, traceable information. Such an approach would not only improve compatibility assessments but also enhance reproducibility, enabling users to track and understand the evolution of models over time.

\subsubsection{How the Right ``Version" Can Be Determined for a New Model}
Once a complete representation of semantic versioning for pre-trained language models is determined, determining the appropriate version ``number" for a new model version requires evaluating how its changes impact backward compatibility and performance. Current practices, as observed in our study, lack systematic tools for this evaluation, resulting in inconsistent version tagging. For example, on Hugging Face, models fine-tuned on new datasets are often assigned major version identifiers, which contradicts the semantic versioning principle that major version increments (X) should be reserved for backward-incompatible changes.  In contrast, fine-tuning a model with a new dataset typically results in a less significant change than altering the base model itself, which may not warrant a major version increment. Instead, a minor version update (Y) is more appropriate in these scenarios. This inconsistency highlights the need for a more structured and standardized approach to version generation.

Building on the concept of semantic version calculators  \citep{semantic_2025}, \citep{lam2020putting} proposed using contracts as inputs for these tools. Extending this idea to PTLMs could involve incorporating model configuration changes, model architecture updates, performance variations, dataset modifications, and dependency changes into the contract. Such an adaptation would provide a framework more suited to the unique needs of PTLMs. Without dedicated tools, developers often rely on intuition or ad hoc practices, which may not fully capture the different changes. Model registries such as Hugging Face could benefit from integrating semantic version calculators to foster consistency, encourage the adoption of community-driven standards, and support users in making the final decision about the right version number. This would enhance transparency and build trust in the evolution of PTLMs.

\subsubsection{How Model Version Metadata can be accessed}
Currently, Hugging Face lacks dedicated fields to store versioning metadata, which forces developers to embed such details in model names. The absence of standardized naming conventions contributes to inconsistencies, making it challenging for users to interpret model changes systematically. While names often include attributes like model size, base model name, and training mechanism, the lack of uniformity results in overloading model names with information, compromising clarity and usability. Short-term efforts should prioritize establishing standard naming guidelines to mitigate these issues and promote consistency across repositories.

In the long term, repositories must address this limitation by introducing structured metadata fields for versioning. These fields should explicitly capture version numbers (e.g., X.Y.Z or a future multi-dimensional representation), compatibility information, and key attributes such as model architecture or task alignment. Decoupling versioning information from model names would establish a more robust framework for tracking changes and ensuring reproducibility. Moreover, incorporating provenance information within model cards or as a separate metadata field would further enhance transparency and accountability in version management, aligning with best practices observed in other domains, such as software supply chain management (e.g., SBOMs).

Therefore, based on our results, we believe the minimal set of essential segments to include in standardized naming or in multidimensional version representations should encompass identifiers, base model information, model size, and training mechanism. As a result, we emphasize the need to standardize naming practices in the short term, and advocate for long-term investments in accessible version metadata. In addition to the need for semantic versioning calculators, future research could explore integrating software bill of materials (SBOM)-inspired tools to decouple versioning and provenance information. SBOM is a formal machine-readable inventory of the components (and their dependency relationships) used for producing a software product \citep{xia2023empirical}. By redefining existing software compatibility notions for PTLMs and establishing robust standards, the ML community can enhance the usability, reproducibility, and transparency of its models. Addressing these challenges requires collaborative efforts between model developers, repository maintainers, and the broader research community, ensuring that PTLM versioning evolves alongside advancements in ML technology.

\subsection{Implications}
Different stakeholders can benefit from our work:

\noindent \textbf{Practitioners:}
\begin{itemize}[label = $\bullet$, labelsep=1em]
    \item Adopt a consistent and structured naming convention that includes segments for base model, size, and version identifiers as a short-term solution to effectively encode versioning information. However, for the longer term, implement explicit versioning mechanisms to provide a more robust and standardized approach to model versioning.
    \item Align versioning practices with the actual development changes of the models and the current stage of model versions to avoid discrepancies between version labels and the true state of the model’s development.
    \item Specify variant type information consistently in model names or descriptions.
    \item Ensure comprehensive training data documentation including dataset sources and preprocessing steps.
    \item Maintain deprecated model tags instead of removing them to aid users in understanding model evolution and informed upgrade decisions.
\end{itemize}

\noindent \textbf{Model registry administrators/Operators:}
\begin{itemize}[label = $\bullet$, labelsep=1em]
    \item Develop and enforce standardized guidelines for naming and versioning models, promote semantic versioning, and provide tools for effective version management.
    \item Establish guidelines for disclosing variant types in model names or descriptions.
    \item Enforce inclusion of detailed training data information in model card submissions.
    \item Implement a mechanism that prevents practitioners from unpublishing or deleting an already published model.
\end{itemize}

\noindent \textbf{Research Community:}
\begin{itemize}[label = $\bullet$, labelsep=1em]
    \item Collaborate on standards and tools supporting structured naming and consistent versioning.
    \item Study the impact of transparent variant type and training data disclosure on user adoption and model performance.
\end{itemize}

\section{Threats to Validity}\label{ttv}
\subsection{Internal Validity}
We only focus on NLP models in this study, although we acknowledge the existence of other models such as Computer Vision, Multimodal, Audio, Tabular, and Reinforcement Learning models. Our focus on language models introduce a threat to internal validity by potentially limiting the generalizability of our findings across different types of models.

To increase internal validity we performed meticulous sampling methods, mitigation of potential threats, and a double-coding (2 people coding in parallel) approach for categorization. Stratified random sampling with a 95\% confidence level and 5\% error rate was employed to select distinct model subsets for each research question (RQ). This approach minimized sampling bias and ensured our findings reflect the diverse range of models relevant to each analysis.

To mitigate potential bias during categorization, a crucial step in all RQs, we implemented a double-coding approach. Two independent researchers systematically interpreted and labeled model names using both open and closed card sorting methods. Initially, open-card sorting was used to identify all relevant terms from \HF, followed by closed-card sorting based on these terms. Any discrepancies in categorization between the coders were resolved through negotiated agreement. Our inter-rater reliability for the open card coding, assessed using Cohen's Kappa, yielded scores indicative of high levels of consistency across all RQs (specific scores mentioned in the relevant sections).

Missing configuration files were handled by using the config.values() function from the \HF Transformers library, ensuring consistent data extraction across all model releases. Additionally, statistically significant sample sizes helped to minimize the influence of random chance on our findings. Through these rigorous methods, we ensured a high degree of internal validity in our study.

Furthermore, the final dataset only includes models with at least 1 million parameters, a threshold we chose to reduce noise and focus on more capable and widely used models for NLP based on \citep{eldan2023tinystories}. This threshold was automatically satisfied by the earlier filtering step removing models without identifiable base models, as the latter models were less likely to meet the size and complexity criteria we aimed for. While this approach helps refine the dataset, there is no universally accepted definition of 'toy models. As a result, some experimental or toy-like models may still be included, as a model with an identifiable base model and more than 1 million parameters might still be fine-tuned or adapted as a toy project or experiment. This presents a potential threat to the internal validity of our study, as the presence of such models could introduce variability in the dataset that might not reflect the characteristics of widely adopted models.

\subsection{External Validity}
Our study aimed to achieve external validity by examining a broader spectrum of model types and naming conventions present on \HF. We carefully documented our methodologies and provided links to specific models, making it easier for other researchers to replicate and generalize our findings. This transparency strengthens the stability of our conclusions within the specific context of model naming practices on \HF. However, we acknowledge several key limitations. First, our findings are based on a specific model registry platform, \HF, which primarily focuses on open-source models. As a result, commercially licensed models might not be represented on the platform. Consequently, the practices observed here might not be fully applicable to the broader \llm landscape. This limitation introduces the potential for selection bias, as \HF's open-source emphasis could differ from the conventions used by other platforms or in proprietary settings.

Second, as discussed in the implications section, the distribution of model usage on HF is skewed, with a small subset of models receiving the majority of attention and engagement. Due to lack of direct information about model usage, our conclusions may vary for models with higher adoption compared to those with limited use. For instance, high-impact models may follow entirely different naming or versioning conventions than less popular ones, and these differences could have a significant impact on the ecosystem. This skewness suggests that future research should investigate how versioning practices and their consequences differ between these groups to provide more understanding of their implications.

Another threat to external validity arises from our reliance on models explicitly specifying their base model as the first step, followed by two heuristics to identify model sizing information: (1) based on safetensors' metadata and (2) explicit size information in the model names. Models that specify sizing information through methods beyond our heuristics may have been incorrectly excluded by our criteria, leading to potential false positives. To mitigate this threat, we based our heuristics on typical cases of model size information observed in the collected data, ensuring that our filter aligns with prevalent practices. Furthermore, our filter is conservative, as only 1.5\% of 53,027 models were filtered out.

To address these limitations, we employed several strategies. First, we acknowledged that the naming and versioning conventions on \HF differ significantly from those on other platforms such as Model Zoo, PyTorch, and ONNX. For instance, on \HF, model names typically follow a two-component structure: owner/model\_name, which clearly indicates ownership and source. In contrast, platforms like Model Zoo, PyTorch, and ONNX use a simpler naming convention, often with a single-component model\_name, without an explicit owner identifier. These differences affect how models are tracked, managed, and versioned across platforms.

To mitigate these platform-specific limitations, we focused on analyzing the underlying principles of versioning and release practices that can be adapted across various model registry platforms, despite their implementation differences. By focusing on these foundational practices, we aimed to derive insights that could be relevant and applicable beyond the specific conventions used by \HF. Furthermore, we encourage the replication of similar analyses on different platforms to enhance the generalizability of our findings and to account for platform-specific nuances in model naming and versioning practices.

\subsection{Construct Validity}
We ensured construct validity through the careful definition and operationalization of key constructs, such as model naming practices and versioning conventions. Our methodologies, including regex-based version extraction and systematic card sorting, align with established practices in software engineering research. The use of Cohen's Kappa to measure inter-rater reliability yielded a score between 0.74 and 0.98, indicating substantial agreement in our labeling process. These measures help ensure that our interpretations accurately reflect the nuances and meanings embedded in model names and versioning practices observed on \HF.

However, potential threats to construct validity may also arise from inconsistencies in data sources or interpretation. To address these, we employed rigorous operational definitions for key constructs, such as reproducibility, variant types, and dataset transparency. Despite these precautions, our study remains dependent on the accuracy and accessibility of data from the \HF API and repository statuses, which could affect the completeness and reliability of our findings.

Construct validity threats may arise from using file size rather than checksums to compare model weight files and determine changes between versions. While file size offers a straightforward comparison method, it may not be sensitive to minor modifications, such as typos, which could still impact the model's behavior but not necessarily alter its size significantly. Checksums, on the other hand, provide a more granular and accurate measure of changes, detecting even small alterations. However, the decision to use file size was influenced by scalability concerns, as handling large model files with checksums could be computationally intensive and time-consuming. Therefore, while file size is a practical choice for scalability, it may introduce potential threats to construct validity due to its limitations in detecting subtle changes.

Another potential threat to construct validity arises from missing predecessors. Some predecessor models were not found in the owner's repository, potentially having been deleted or removed. This absence could affect the completeness of our analysis and the accuracy of our measurements of model changes. By not replacing missing predecessors with alternative models, we aimed to maintain the integrity of our dataset and avoid introducing potential biases. However, this decision may impact the representation of versioning practices and model evolution, potentially affecting the validity of our findings.

\section{Conclusion}\label{conclusion}
In this study, we conducted a comprehensive investigation into \llm releases on HF, focusing on naming and versioning conventions, release transparency, and differences between major and minor versions. Utilizing a mixed-method approach combining quantitative and qualitative analyses, we provided nuanced insights into the landscape of \llm releases on \HF. Our study addressed three primary research questions: the naming and versioning conventions of \llms on \HF, the provenance, transparency, and reproducibility of \llm releases, and the differences between major and minor versions.

We found 148 naming practices for \llms on \HF, characterized by segment counts and semantic meanings. We identified major and minor versioning patterns, indicating significant updates and incremental changes, respectively. Notably, 98\% of \llms included configuration files, enhancing reproducibility and transparency. However, 29.28\% \llms explicitly mentioned \vtypes and included references to training datasets, limiting transparency and reproducibility.

Through manual and statistical analyses, we observed significant differences between major and minor predecessor-successor pairs. Major updates involved substantial changes in \bases and model weight files, with a total of 28 unique changes. In contrast, minor updates exhibited incremental modifications, with a total of 8 unique changes. These minor updates often overlapped with the changes in the major versions. Additionally, 524,419 version traces were embedded in commits without being indicated in the model names or repository, highlighting the need for semantic versioning on \HF. Building on these findings, there are opportunities for future work to improve the release process of \llms. 

Future research should focus on establishing semantic versioning practices for \llms. 

\section*{Data Availability}
\label{sec:availability}
The datasets generated and analyzed during this study are available in the replication package~\citep{SAILResearch2024}.
\section*{Funding} 
This research was supported by the NSERC Discovery Grant.
\section*{Ethical Approval} This study does not involve human participants or animals.
\section*{Informed Consent} No human subjects were involved in this study.
\section*{Conflicts of Interests/Competing Interests}
The authors declare that they have no known competing interests or personal relationships that could have (appeared to) influenced the work reported in this article.
\section*{Author Contributions}
\begin{itemize}
    \item Adekunle Ajibode: Conceptualization, Data Collection, Methodology, Data Analysis, Writing – Original Draft.
    \item Abdul Ali Bangash: Methodology, Data Validation, Writing – Review \& Editing.
    \item Filipe Roseiro Cogo: Methodology, Writing – Review \& Editing.
    \item Bram Adams: Supervision, Writing – Review \& Editing, Conceptual Guidance, Research Direction.
    \item Ahmed E. Hassan: Supervision, Research Direction.
\end{itemize}:

\bibliographystyle{plainnat}
\bibliography{bibliography}
\end{document}